\documentclass[]{article}

\usepackage[margin=1in]{geometry}

\usepackage{graphicx}

\usepackage{amsmath}
\usepackage{rotating}
\usepackage{amssymb}
\usepackage{algorithm}
\usepackage{algorithmic}

\usepackage{xcolor}
\usepackage{subcaption}
\usepackage{mathtools}
\usepackage{color}
\usepackage{url}

\usepackage{listings}
\graphicspath{ {./images/} }

\newcommand{\Prod}[2]{\left({#1}, {#2}\right)}

\author{Pawe\l{} Maczuga$^1$, Marcin \L{}o\'s$^1$, Eirik Valseth$^{2,3,4}$, Albert Oliver Serra$^5$, \\ Leszek Siwik$^1$,  Elisabede Alberdi Celaya$^6$, Anna Paszy\'nska$^7$,  Maciej Paszy\'nski$^1$}

\date{$^1$ AGH University of Krakow, Poland \\ $^2$ Norwegian University of Life Sciences, Norway \\ 
$^3$ The Oden Institute for Computational Engineering and Sciences \\ at The University of Texas at Austin,  Texas, USA   \\ $^4$ Simula Research Laboratory, Norway \\
$^5$ University of Las Palmas de Gran Canaria, Spain \\ 
 $^6$ The University of the Basque Country, Bilbao, Spain \\
$^7$ Jagiellonian University, Krak\'ow, Poland }

\begin{document}
\title{Simulating the aftermath of \\
Northern European Enclosure Dam (NEED) break \\ and flooding of European coast}

\maketitle

\begin{abstract}
The Northern European Enclosure Dam (NEED) is a hypothetical project to prevent flooding in European countries following the rising ocean level due to melting arctic glaciers. This project involves the construction of two large dams between Scotland and Norway, as well as England and France. The anticipated cost of this project is 250 to 500 billion euros. In this paper, we present the simulation of the aftermath of flooding on the European coastline caused by a catastrophic break of this hypothetical dam. 
From our simulation results, we can observe that there is a traveling wave after the accident, with a velocity of around 10 kilometers per hour, raising the sea level permanently inside the dammed region. This observation implies a need to construct additional dams or barriers protecting the northern coastline of the Netherlands and the interior of the Baltic Sea. Our simulations have been obtained using the following building blocks. First, a graph transformation model was applied to generate an adaptive mesh approximating the topography of the Earth. We employ the composition graph grammar model for breaking triangular elements in the mesh without the generation of hanging nodes. Second, the wave equation is formulated in a spherical latitude-longitude system of coordinates and solved by a high-order time integration scheme using the generalized $\alpha$ method. 
\end{abstract}

\section{Introduction}

Rising ocean levels caused by climate change and melting Arctic glaciers have worried European environmentalists to the point that a plan has been made to build a large dam to separate the oceans from the North Sea and Baltic Sea basins. The Northern European Enclosure Dam (NEED) project \cite{NEED2,NEED3}  contemplates building two huge dams, the first between Scotland and Norway and the second between England and France. This would require the construction of four sections of dams in the north and one section in southern England. The first dam, 331 kilometers long respectively, would connect the west coast of Norway with the Scottish island of Shetland. Another section between the Isle of Shetland and Scotland with a total length of 145 kilometers, and another covering the Orkney islands. In addition, a dam would need to be built in the south of England between England's Cornwall and Bretagne in France, with a length of 161 kilometers.

The water depth at the dam sites reaches 110 meters at its deepest points. Technical details of the construction of the NEED dam are under consideration, but nevertheless, preliminary estimates have put the cost of building such a dam at an expense of 250-500 billion euros. Despite such an extreme cost, the project is being considered very seriously since the Arctic glaciers are melting \cite{NEED2}. 
Since 1880, the average sea level has risen by 21 centimeters \cite{Church}, and the predicted average increase in ocean levels is 2.3 meters for every one degree Celsius increase in average temperature \cite{Levermann}. 
The result, as noted by \cite{NEED4}, is an inevitable rise in sea levels in the range of 5-11 meters over the next centuries. 
The bleakest scenarios assume a more than 10-meter rise in global average ocean levels by the year 2500 \cite{DeConto,Edwards}. 
More cautious scenarios described in \cite{Jevrejeva,Kopp,Bars} predict a 1-2 meter rise in average ocean levels by 2100.
A recent paper \cite{Arctic} shows that the West Antarctic Ice Sheet encountered rapid melting 8000 years ago, which could cause an increase of the average ocean level up to 4 meters in 200 years. 

We focus on a hypothetical scenario where we assume that the Northern Sea and Atlantic Ocean levels rise by 6 meters and the Northern and Baltic Seas are protected by the NEED. This is the starting point of our consideration. At this moment, we assume a massive dam break. There may be several potential causes of such a scenario, including natural causes like an earthquake or asteroid strike, as well as material fatigue or design flaws. Another theoretically possible scenario is a terrorist attack against the dam. 

To perform our simulation, we employ the following tools:
\begin{itemize}
\item A computational mesh is generated using a novel composition graph grammar model described in Section \ref{sec:CPgraph}. The topography of the Earth and the seabed is based on the Global Multi-Resolution Topography Data Synthesis database \cite{GMRT}.
\item A simulation that employs the wave equations summarized in Section \ref{sec:wave} discretized in space using finite element method over the entire Earth mesh.
\item A time integration scheme which employs a high-order generalized $\alpha$ method described in Section \ref{sec:genalpha}.
\end{itemize}
The simulation of the NEED dam break using all the above tools is summarized in Section
\ref{sec:break}.
In Section \ref{sec:parallel} we also discuss the parallel scalability of our implementation. The paper is concluded in Section 
\ref{sec:conclusions}.

\section{{The first building block, a simple composition graph grammar model for breaking triangular elements in the mesh without generation of hanging nodes}}
\label{sec:CPgraph}

{The first building block for our simulator is the composition graph grammar model. It models Rivara's algorithm \cite{Rivara1,Rivara2} for breaking triangular elements in a way that does not generate hanging nodes. 
All the triangles of the mesh are represented by the composition graph grammar \cite{FIAnia}. The way we break the triangular elements is described by some rules that transform the composition graph. These rules are defined following the idea of the composition graph grammar proposed by Ewa Grabska \cite{Grabska1,Grabska2}.
The formal definition of the composition graph grammar can be found in
\cite{FIAnia}.}

{The Rivara longest-edge refinement algorithm is summarized in Figure \ref{fig:Rivara}. 
The main assumption of the Rivara algorithm is that we can only break the longest edges of elements. This prevents the creation of elongated elements that may cause numerical problems.
We start with four triangular finite element meshes. We want to break the first triangle; see panel (a) in Figure \ref{fig:Rivara}. We mark its longest edge by the red dot; see panel (a) in Figure \ref{fig:Rivara}. The general condition says that \emph{we can break the marked edge only if two neighboring triangles share it, and for both triangles, this edge is the longest. Another option is when this edge is located on the boundary.}
This is not the case here since the second element has another longest edge. Thus, we mark the other longest edge of the second element, which we denote again by the red dot, see panel (a) in Figure \ref{fig:Rivara}. This procedure is repeated; see panel (b) in Figure \ref{fig:Rivara} until we end up with the edge located on the side of the domain. This time it can be broken. We break the last element; see panel (b) in Figure \ref{fig:Rivara}. After this operation, the previous red dot is located on the edge that fulfills the breaking condition; see panel (c) in Figure \ref{fig:Rivara}. We can break this edge and the two adjacent elements; see panel (c) in Figure \ref{fig:Rivara}. The same happens to the second edge denoted by the red dot in panel (c) in Figure \ref{fig:Rivara}. We can break this edge. Finally, the first edge denoted by the red dot also fulfills the breaking condition; see panel (c) in Figure \ref{fig:Rivara}. We can break the two adjacent elements, including the first element that was our original target, see panel (c) in Figure \ref{fig:Rivara}.}

{The composition graph grammars have been successfully used to model mesh refinements with triangular or rectangular elements
\cite{CPgraph1,CPgraph2,FI1,FI2}. The definition of the Rivara algorithm by another kind of transformative rules called the hypergraph grammar, has already been described in \cite{EC}. 
There are the following differences between the previous model described in \cite{EC} and the new model presented in this paper:
\begin{itemize}
\item The previous model employs the hypergraph grammar model, while the new model uses the composition graph grammar model. 
\item The previous model needs six hypergraph grammar productions presented in Figures 5-10 in \cite{EC}, while the new model needs only four composition graph grammar productions presented in Figures \ref{fig:P1}-\ref{fig:P4}.
\item The six productions from the hypergraph grammar model described in \cite{EC}, presented in Figures 5-10 there, have complex left-hand sides, which are very costly to implement. The subgraphs of the left-hand sides of these productions need to be identified as subgraphs of the big hypergraph representing the entire computational mesh there. They consist of triangular elements with between 3 to 5 vertices and several hyperedges. The four productions of our new composition graph model presented in Figures \ref{fig:P1}-\ref{fig:P4} have only one or two vertices, so their identification is straightforward.
\end{itemize}
Summing up, our new model presented in Figure \ref{fig:P1}-\ref{fig:P4} is simpler and cheaper to implement than our previous model described in \cite{EC}.}

The Rivara algorithm has also been expressed by graph-grammar in \cite{tsunami5}. There are the following differences between \cite{tsunami5} and this paper:
\begin{itemize}
\item In \cite{tsunami5}, the graph-grammar-based mesh generation has been performed in the spherical system of coordinates, but the PDE has been formulated in the Cartesian system of coordinates, and the simulation was performed only on small part of the domain. In our new paper, the PDEs are formulated in the spherical system of coordinates, and the simulation is performed on the entire mesh.
\item The graph grammar model presented in \cite{tsunami5} expressed the Rivara algorithm by two productions, with four connected vertices and six connected vertices on the left-hand sides. These left-hand sides of the productions are still expensive to identify. In this new paper, we introduce graph-grammar productions that have only one or two vertices on the left-hand sides, and they are easy to identify to execute graph grammar productions.
\item The computations performed in \cite{tsunami5} were sequential, while the computations presented in this paper are parallel.
\end{itemize}

\begin{figure}[!htb]
  \centering
\includegraphics[width=0.7\textwidth]{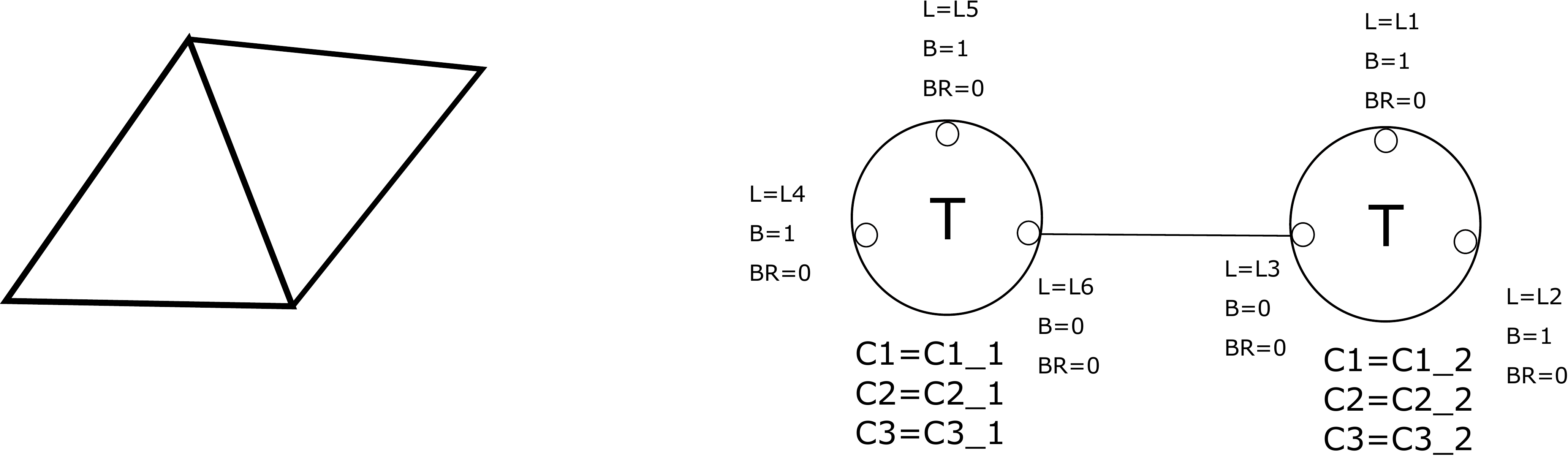}
  \caption {An exemplary two-element mesh and the corresponding CP-graph.}
  \label{fig:2elementmesh}
\end{figure}

\begin{figure}[!htb]
  \centering
\includegraphics[width=0.7\textwidth]{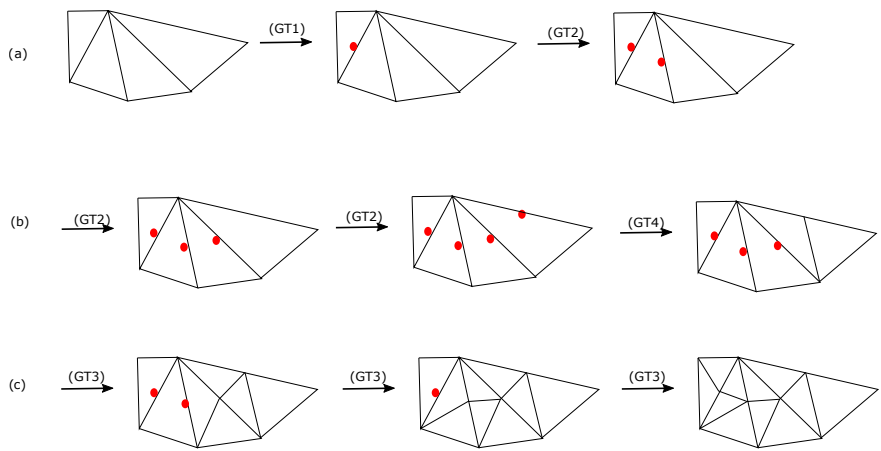}
  \caption{Rivara algorithm of refinement, step by step.}
  \label{fig:Rivara}
\end{figure}

\begin{figure}[!htb]
  \centering
\includegraphics[width=0.7\textwidth]{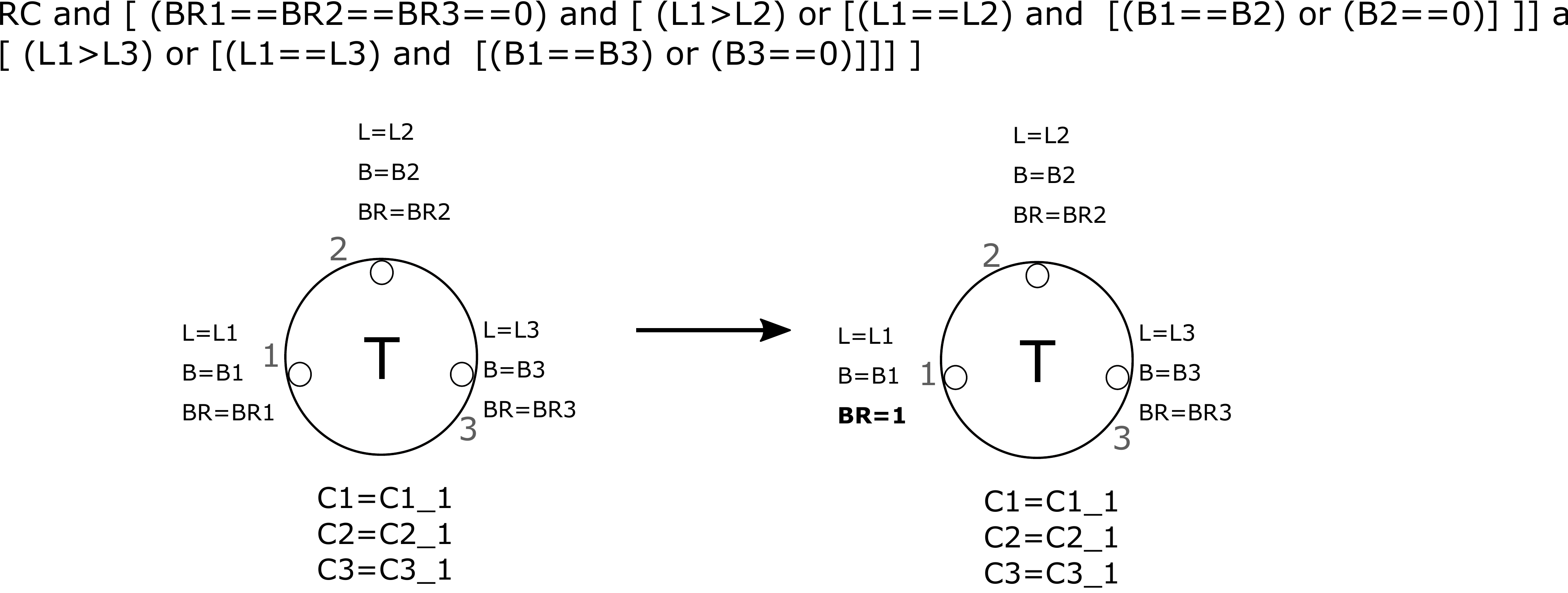}
  \caption{Production $GT1$, marking element for breaking.}
  \label{fig:P1}
\end{figure}

\begin{figure}[!htb]
  \centering
\includegraphics[width=0.9\textwidth]{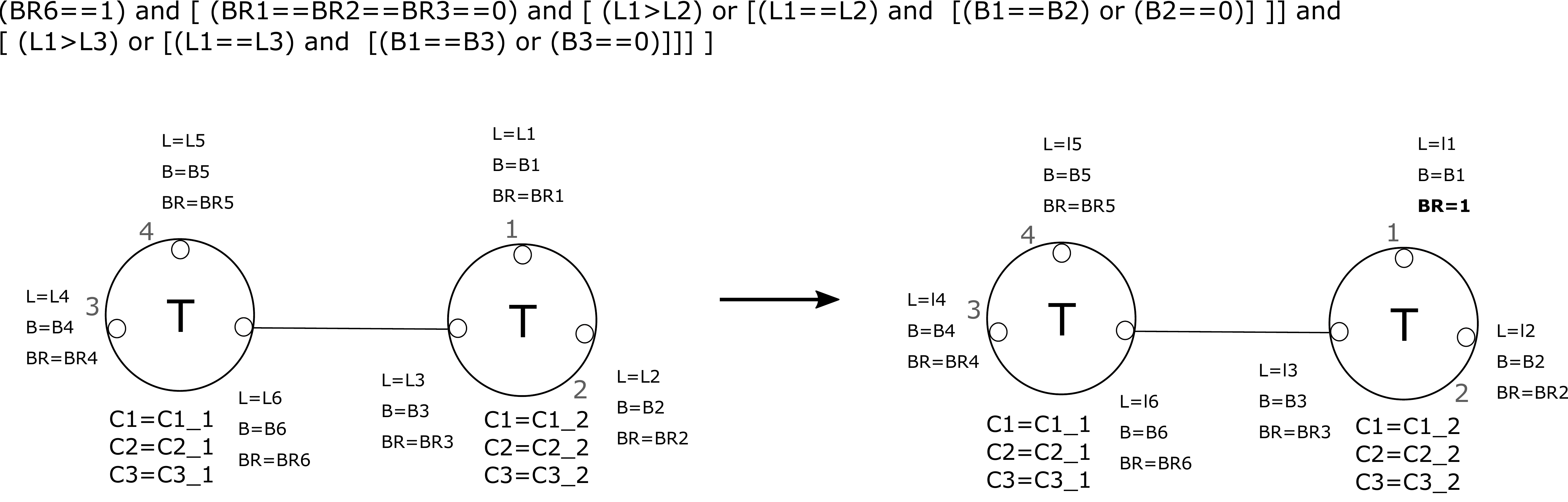}
  \caption{Production $GT2$, propagation of refinement marker.}
  \label{fig:P2}
\end{figure}

\begin{figure}[!htb]
  \centering
\includegraphics[width=1.0\textwidth]{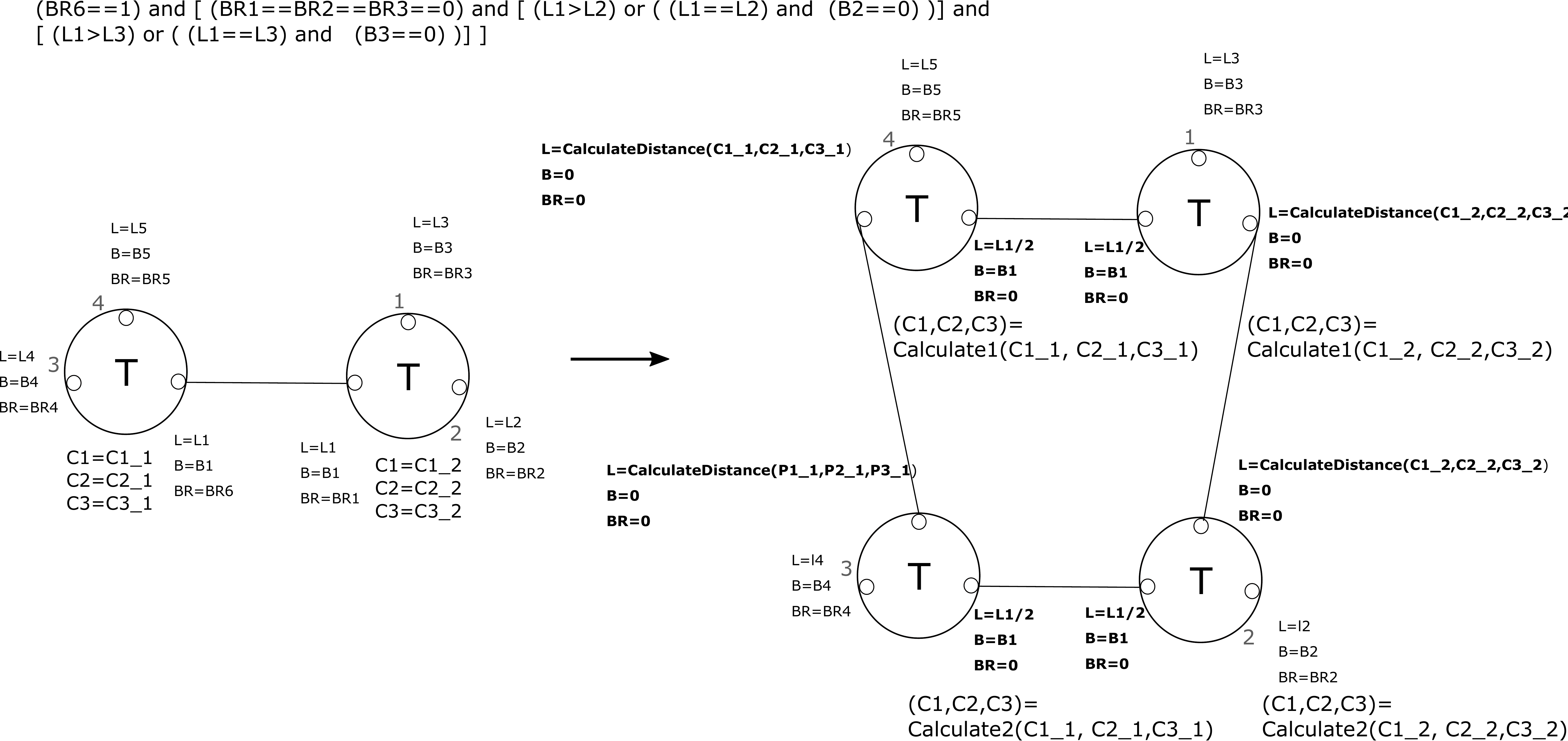}
  \caption{Production $GT3$, breaking of common longest edge.}
  \label{fig:P3}
\end{figure}

\begin{figure}[!htb]
  \centering
\includegraphics[width=0.3\textwidth]{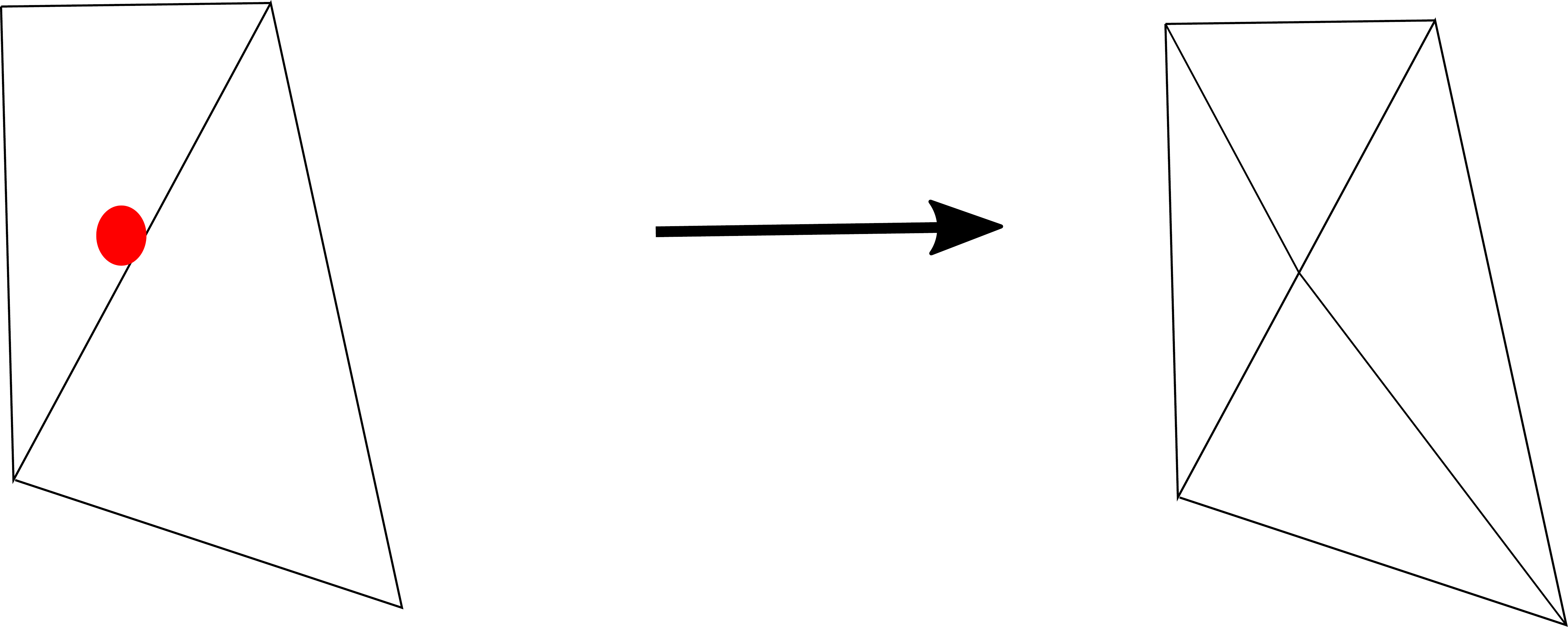}
  \caption{Breaking of two neighboring elements.}
  \label{fig:P3_on_mesh}
\end{figure}

\begin{figure}[!htb]
  \centering
\includegraphics[width=0.8\textwidth]{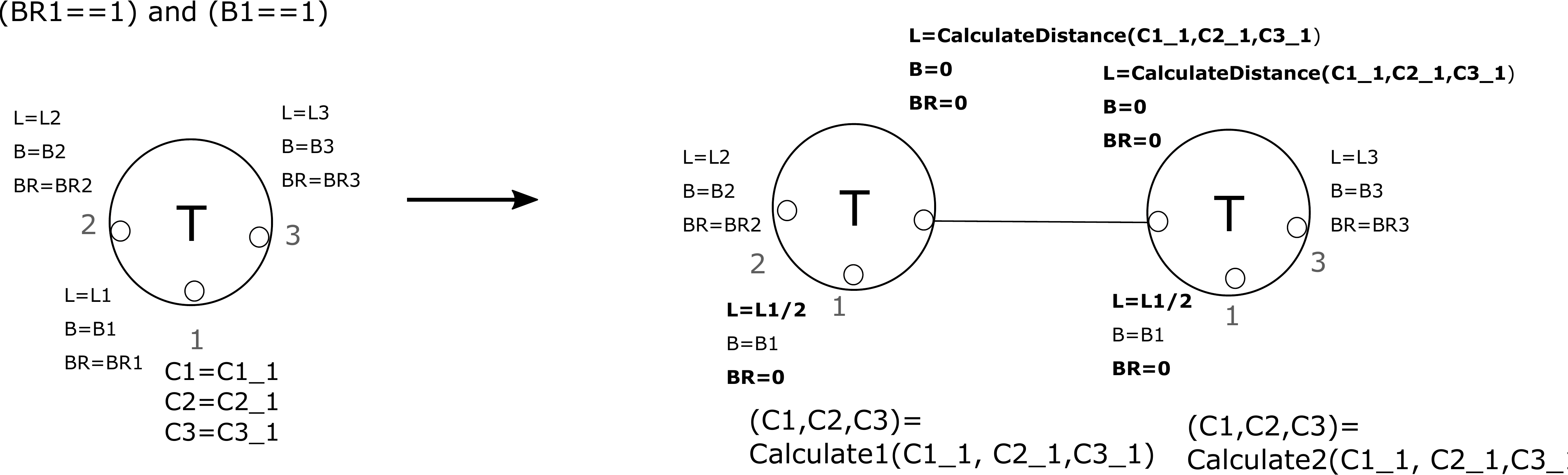}
  \caption{Production $GT4$, breaking of boundary edge.}
  \label{fig:P4}
\end{figure}

{Let us now introduce the new composition graph grammar model for breaking two-dimensional triangular elements without the creation of the hanging nodes.}
Figure \ref{fig:2elementmesh} presents an exemplary two-element mesh and the corresponding CP-graph. A node represents a triangular element. Each node has label $T$ and attributes $C1, C2, C3$ the coordinates of the triangle vertices.
Each node has three bonds representing the edges of a triangle. Each edge has three attributes: attribute $L$ - the length of the edge; attribute $B$, which equals $1$ if the edge is boundary edge and "0" in the other case; and attribute $BR$, which equals $1$ if the edge should be broken and "0" in the other case.

The graph grammar modeling the longest edge Rivara algorithm consists of four productions:
\begin{itemize}
    \item Production (graph transformation) $GT1$ (Figure \ref{fig:P1}) which marks for refinement the longest edge of an element that should be refined (element for which the refinement criteria ($RC$) is fulfilled). If an element has more than one longest edge, the boundary edge will be chosen. The value of the attribute $BR$ for the longest edge is set to $1$.
    \item Production (graph transformation) $GT2$ (Figure \ref{fig:P2}) which propagates the refinement - marks for refinement of the longest edge of the neighboring element. If the neighboring element has more than one longest edge, the boundary edge will be chosen. The value of the attribute $BR$ for the longest edge is set to $1$.
    
    \item Production (graph transformation) $GT3$ (Figure \ref{fig:P3}), which propagates the refinement in the case when the neighboring element's longest edge is the common edge and performs refinement of both elements to the longest common edge. This production replaces the CP-graph representing two big neighboring elements with a CP-graph representing four smaller elements. Figure \ref{fig:P3_on_mesh} presents changes in the two-element mesh when we perform refinement (graph transformation $GT3$).
    The function $Calculate1(C1,C2,C3)$ and $Calculate2(C1,C2,C3)$ calculate the coordinates of vertices of the newly created first and second small triangle, respectively, on the base of coordinates of vertices of the big triangle. The function $CalculateDistance(C1,C2,C3)$ calculates the length of the newly created common edge.
    \item Production (graph transformation) $GT4$, presented in Figure \ref{fig:P4}, performs refinement of the longest edge, which is the boundary edge. This production replaces the CP-graph representing one big boundary element with a boundary edge marked for refinement by a CP-graph representing two smaller elements.
    The function $Calculate1(C1,C2,C3)$ and $Calculate2(C1,C2,C3)$ calculate the coordinates of vertices of the newly created first and second small triangle, respectively, on the base of coordinates of vertices of the big triangle.
    
\end{itemize}


\begin{figure}[!htb]
  \centering
\includegraphics[width=1.1\textwidth]{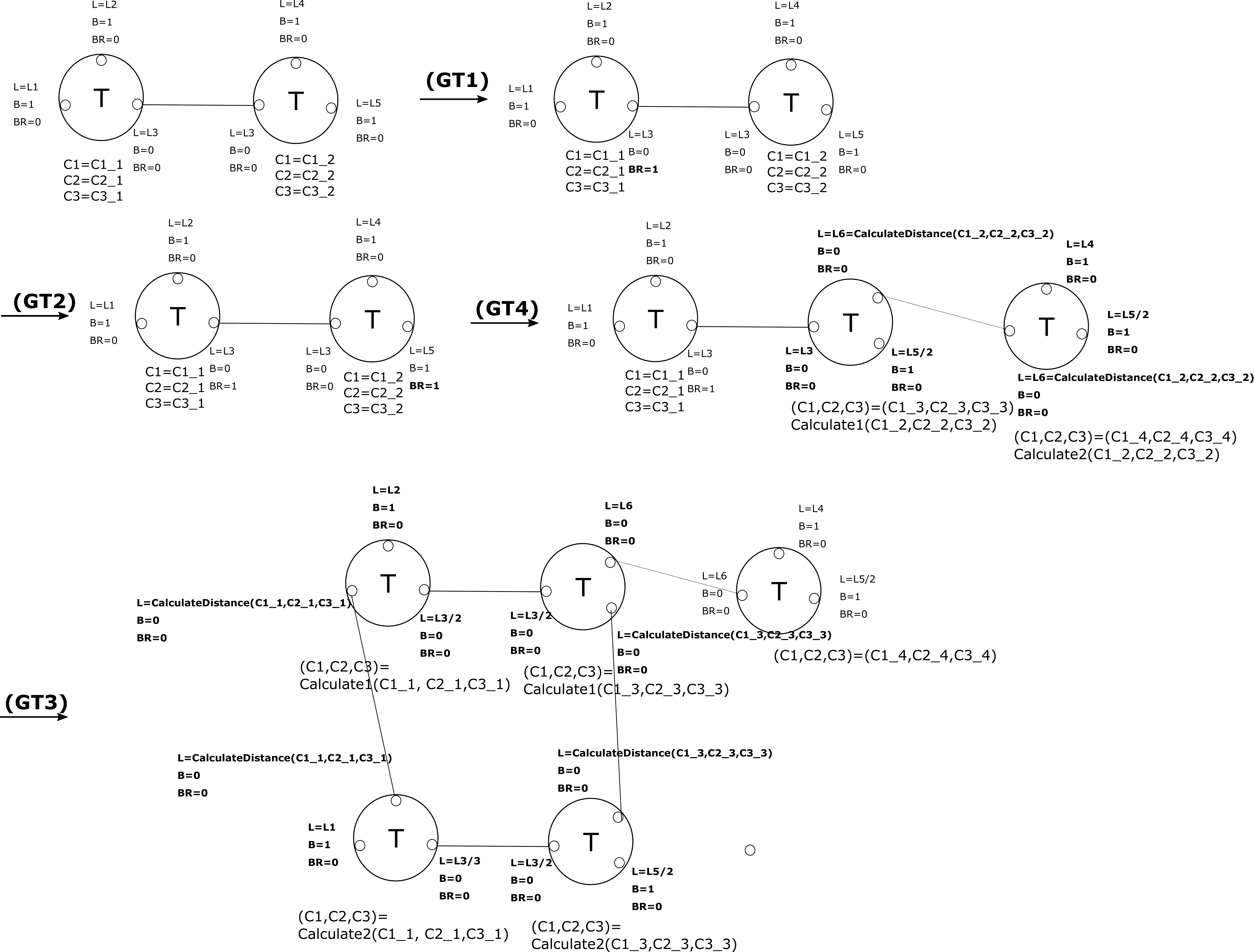}
  \caption{An exemplary derivation modeling the process of Rivara algorithm as presented in Figure \ref{fig:Rivaraexample}.}
  \label{fig:derivation}
\end{figure}

\begin{figure}[!htb]
  \centering
\includegraphics[width=0.8\textwidth]{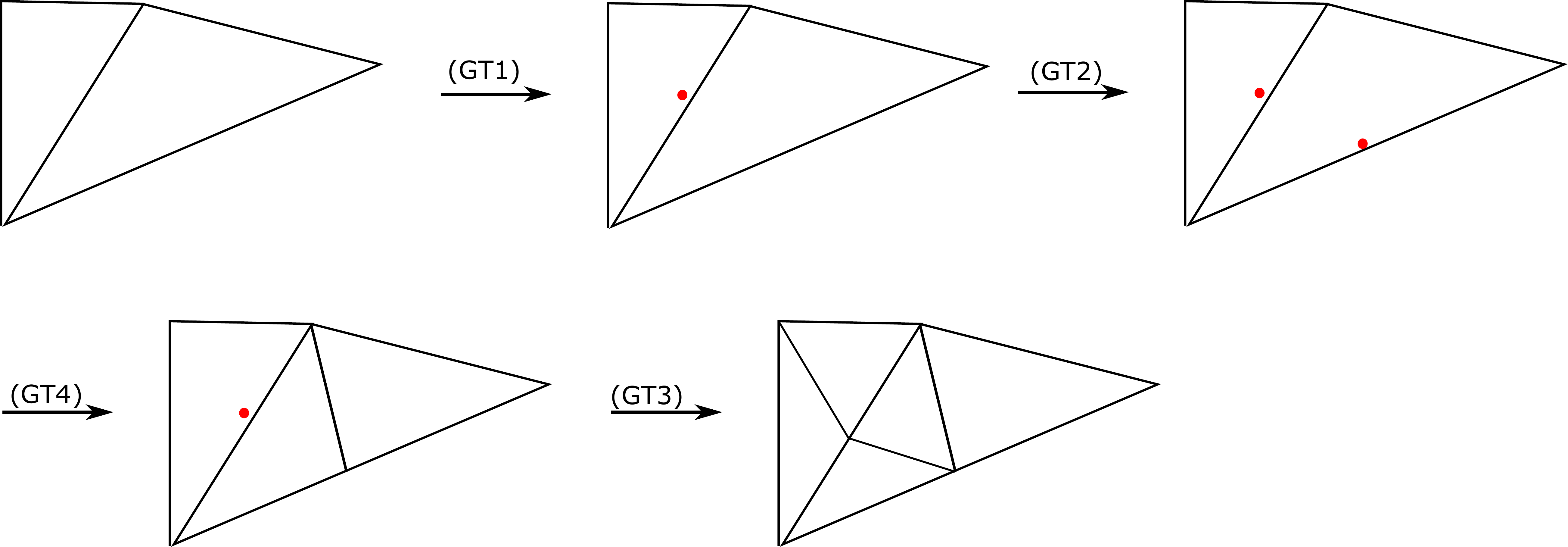}
  \caption{The process of Rivara refinement algorithm for the simple two element mesh in the case the left element should be refined.}
  \label{fig:Rivaraexample}
\end{figure}

Figure \ref{fig:derivation} presents derivation modeling the process of Rivara refinement algorithm as presented in Figure \ref{fig:Rivaraexample}.

\section{The second building block modeling the dam break with the wave equations} \label{sec:tsunami_sim}
\label{sec:wave}

The second building block for our simulations are the wave equations. A comprehensive review of modeling tsunami waves is described in ~\cite{tsunami4}. Seawater flow is often modeled using shallow water theory and leads to a set of transient and nonlinear PDEs, i.e., the shallow water equations (SWE)~\cite{TanShallow1}. 
The solution of the SWE for physically relevant flows 
requires significant resources due to its nonlinear nature and multiple unknowns. Multiple surrogates exist that are valid in certain limited cases and admit accurate solutions within their region of validity. In this case of tsunami modeling, we will focus on the case in which the wave propagation is governed by long wave theory. 

\begin{figure}
  \centering
\includegraphics[width=0.7\textwidth]{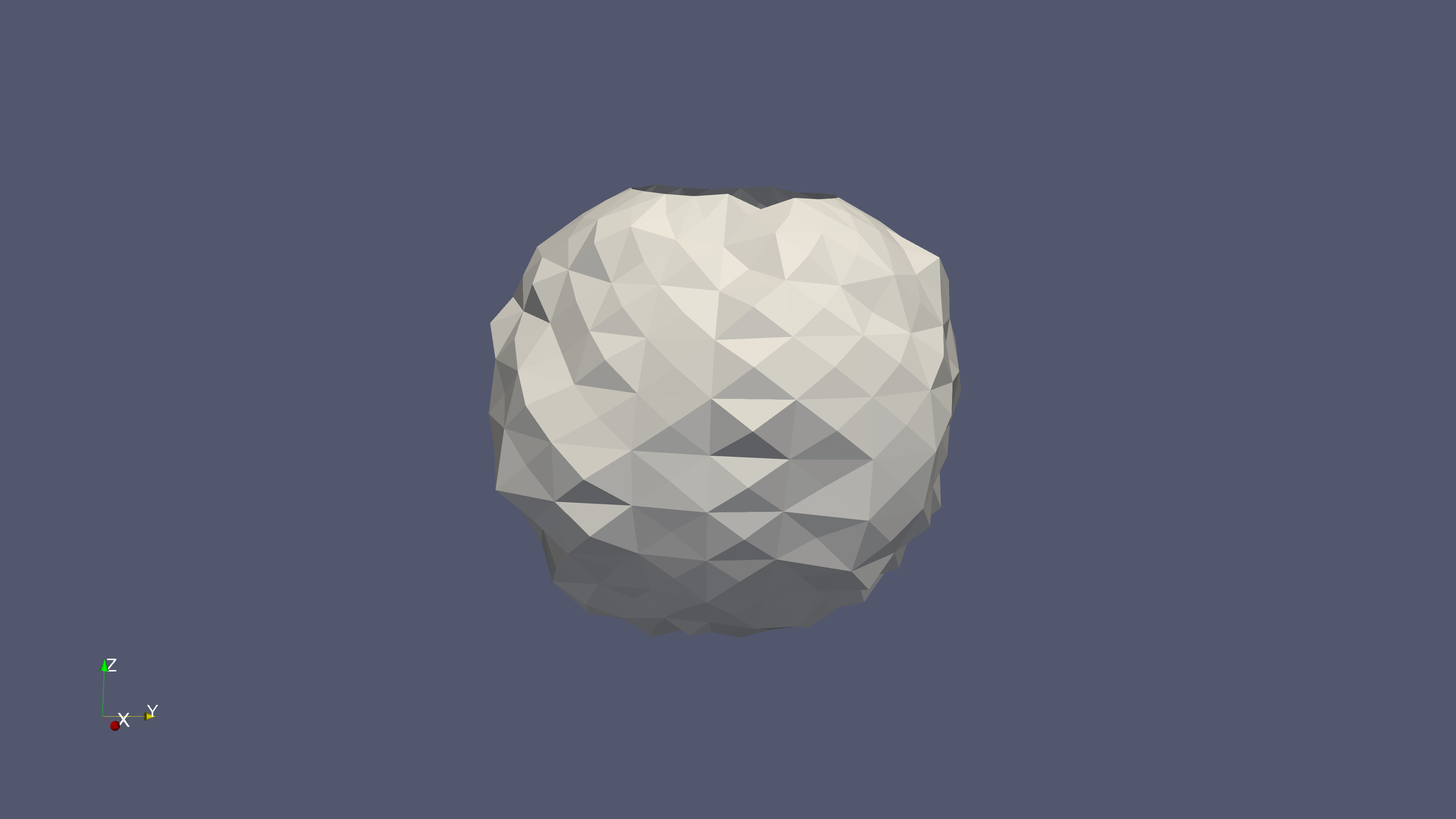}\\ \includegraphics[width=0.7\textwidth] {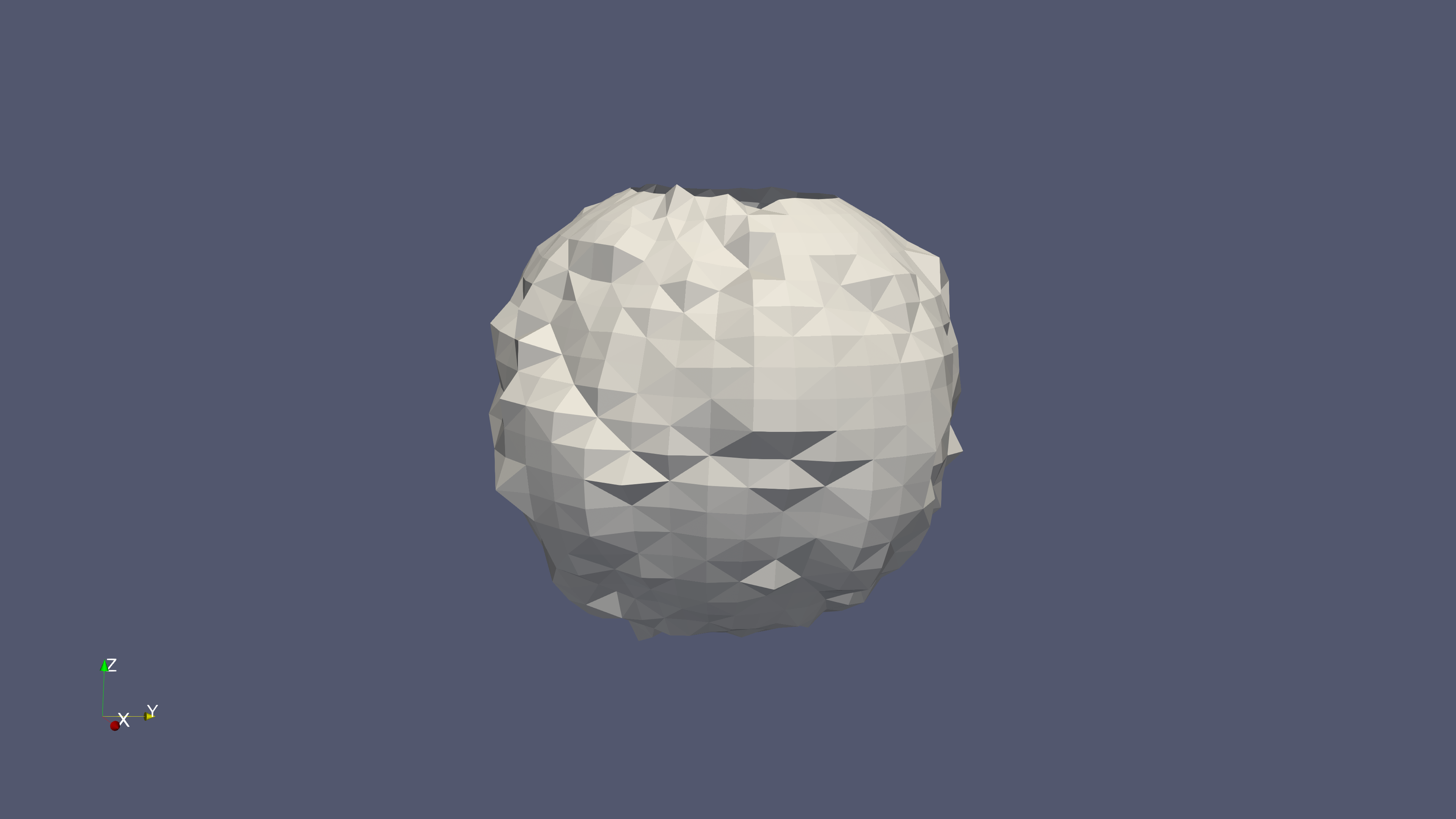}\\ \includegraphics[width=0.7\textwidth] {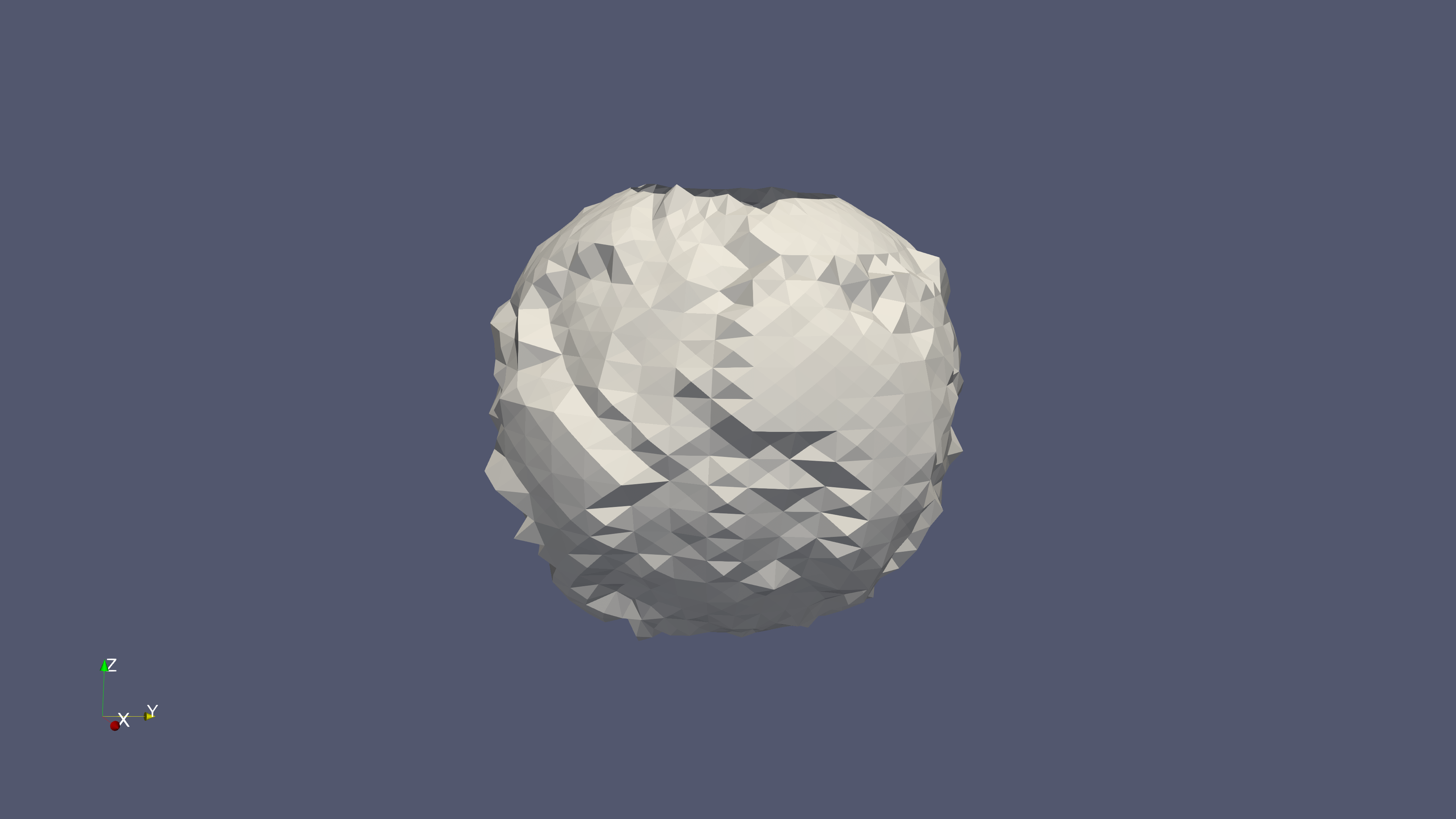} 
\caption{Generation of the topography of the Earth (1/3) with composition graph grammar expressing the Rivara method.}
  \label{fig:Earth1}
\end{figure}

\begin{figure}
  \centering
\includegraphics[width=0.7\textwidth]{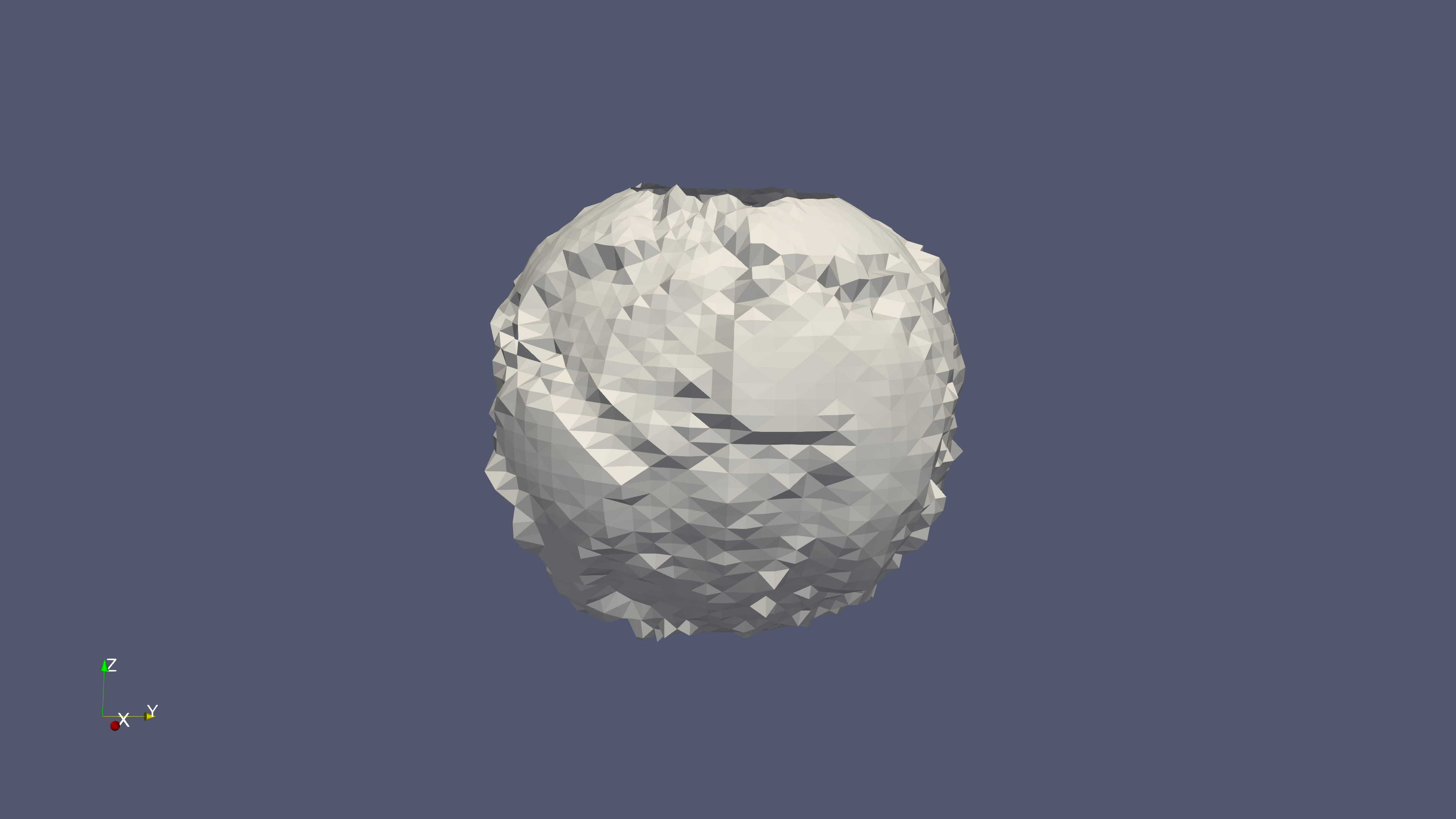}\\
\includegraphics[width=0.7\textwidth]{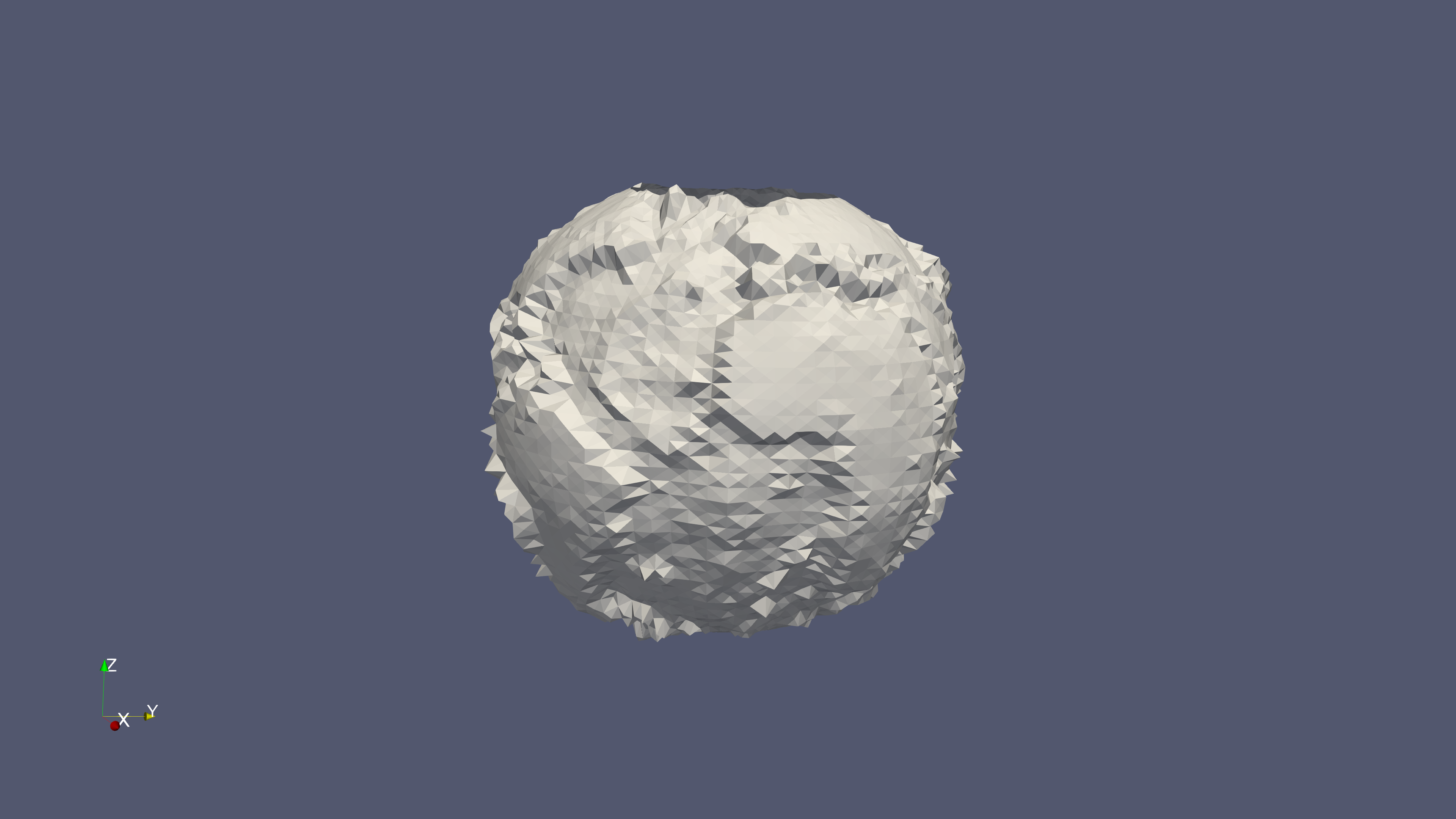}\\ \includegraphics[width=0.7\textwidth]{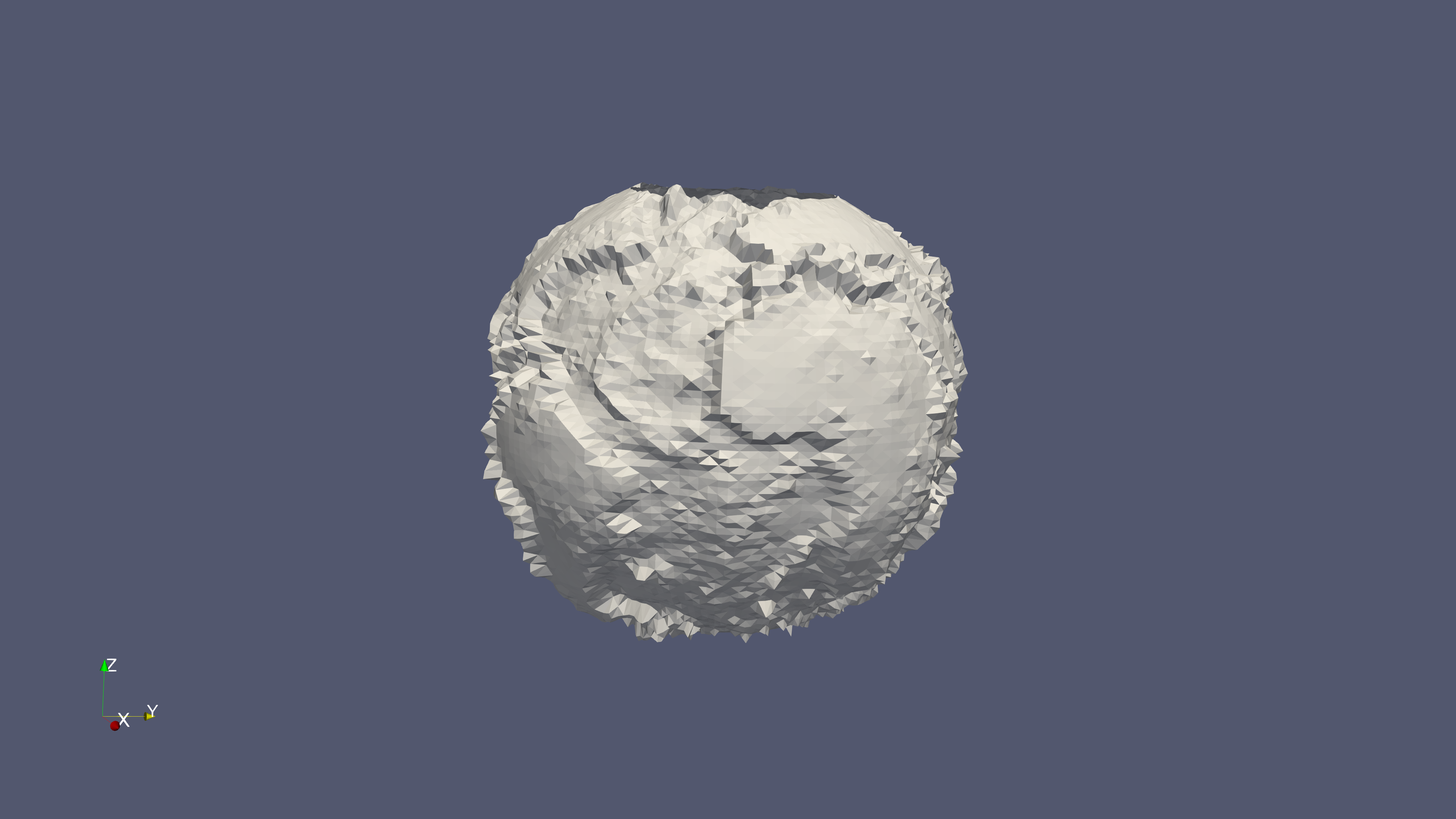}
\caption{Generation of the topography of the Earth (2/3) with composition graph grammar expressing the Rivara method.}
  \label{fig:Earth2}
\end{figure}

\begin{figure}
  \centering
\includegraphics[width=0.7\textwidth]{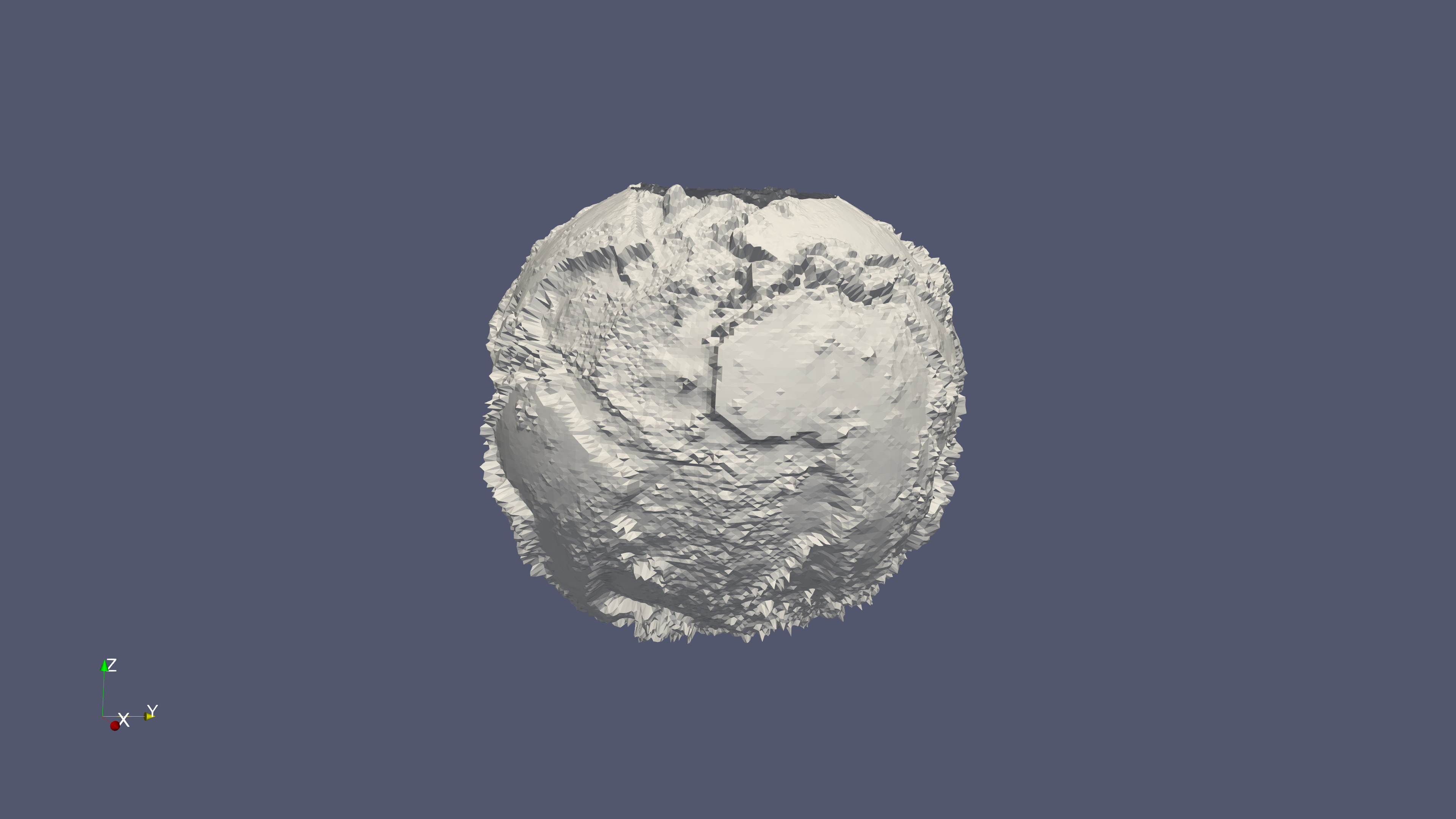}\\
\includegraphics[width=0.7\textwidth]{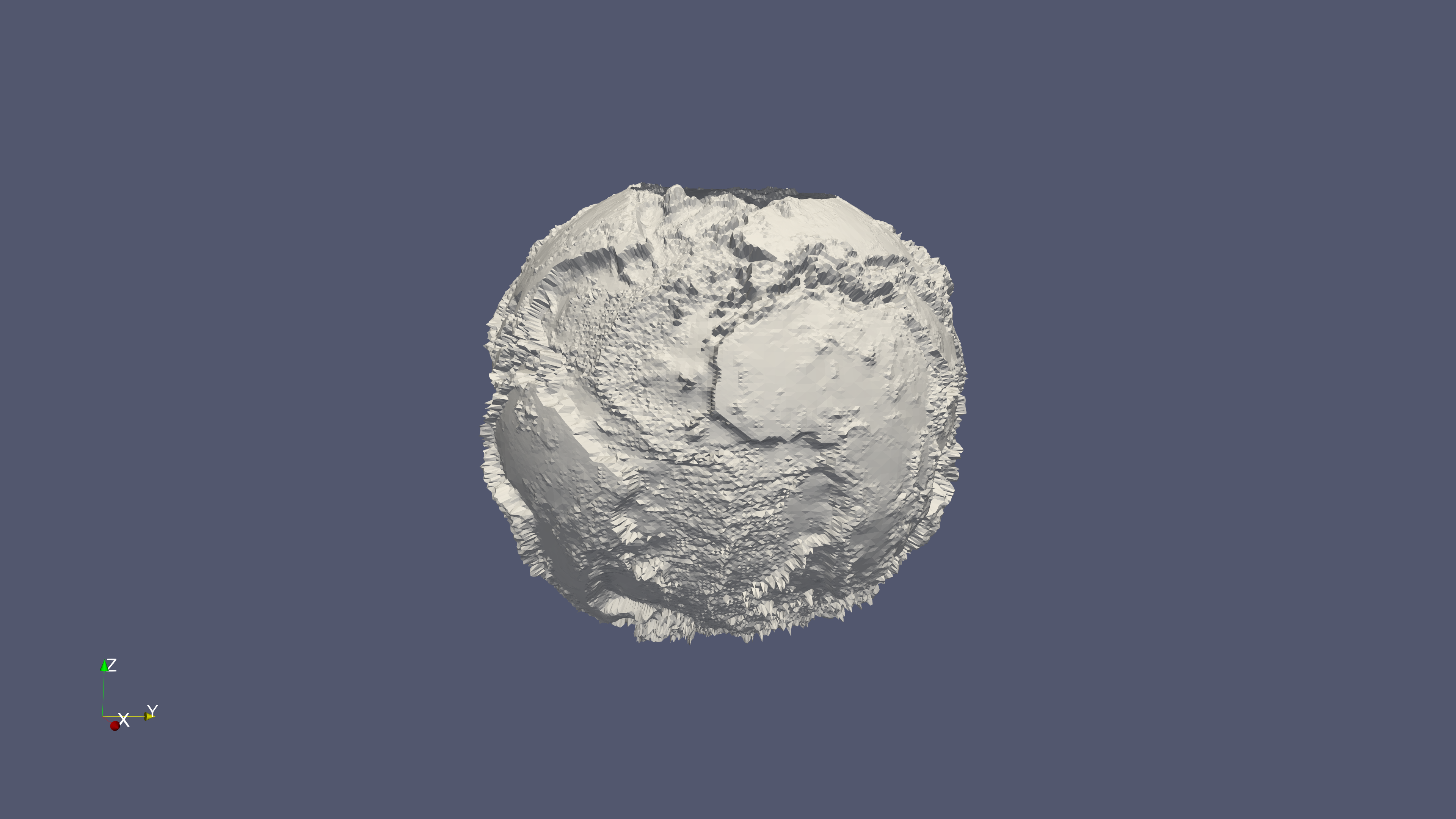}\\ \includegraphics[width=0.7\textwidth]{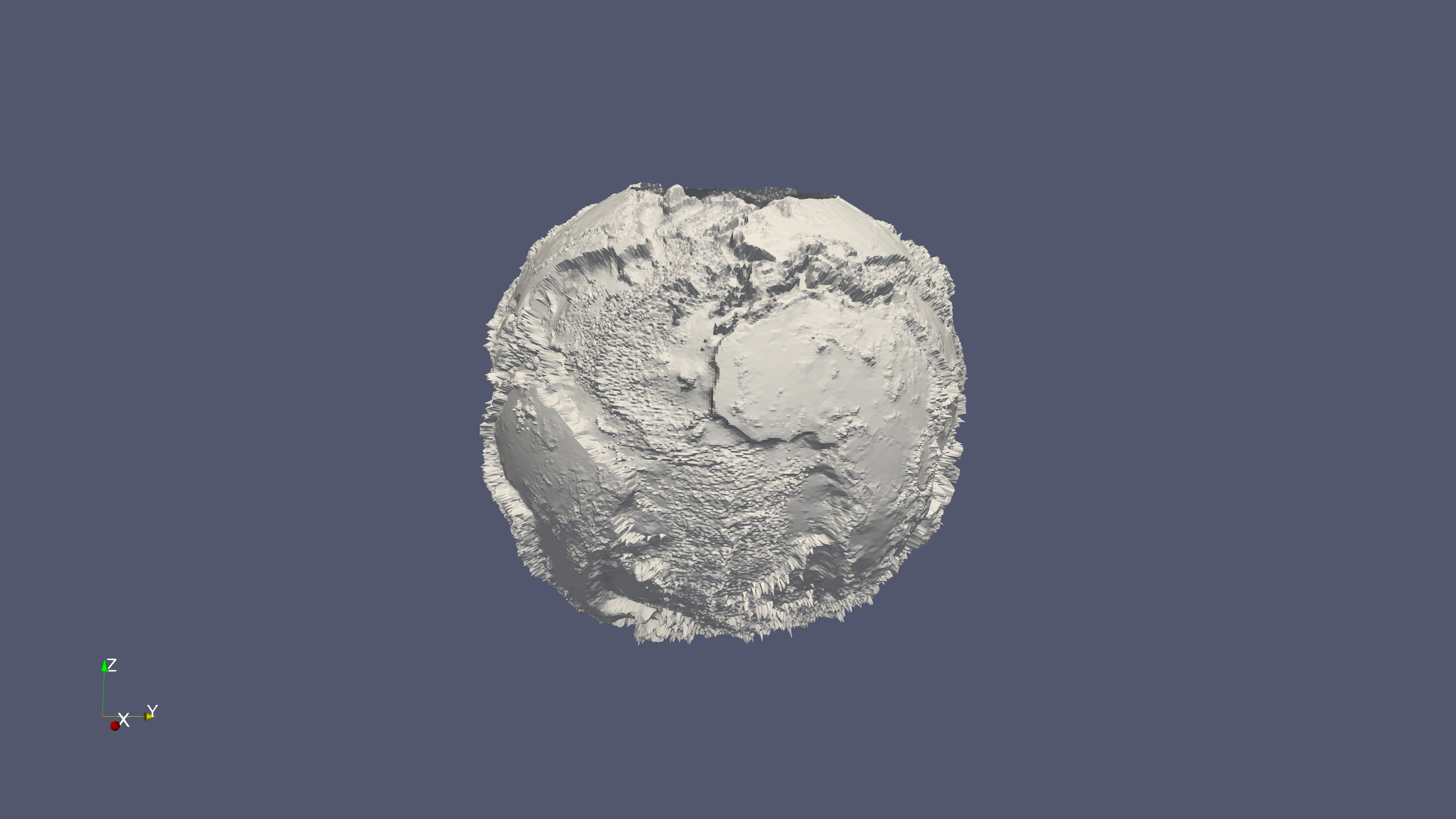}
\caption{Generation of the topography of the Earth (3/3) with composition graph grammar expressing the Rivara method.}
  \label{fig:Earth3}
\end{figure}

The approach we use has been studied extensively and appears in, e.g.,~\cite{tsunami1,tsunami2,tsunami5}. Starting with nonlinear shallow water theory and performing modeling assumptions (see~\cite{tsunami4}), the SWE can be reduced to the following wave equation: 
\begin{eqnarray} \label{eq:wave_eq}
\frac{\partial^2 u}{\partial t^2} - \nabla \left( g(u-h_b) \nabla u \right) = 0 \text{ in } \, \Omega.
\end{eqnarray}
In this formula, the unknown scalar field $u$ represents the water level, thus, $u-h_b$ is the water depth relative to the seabed denoted by the $h_b$. 
The acceleration due to gravity is denoted by  $g=9.81 \frac{m}{s^2}$.
The computations are performed on the model of the whole Earth, where the North Pole and South Pole are modeled as the zero Neumann baoundary conditions.
Here, we have assumed that the wave speed is the shallow water wave celerity, i.e., $c^2 = g(u-h_b)$. Note that~\eqref{eq:wave_eq} is the standard representation of the nonlinear wave equation in  $x,y,z$ coordinates. 
As the computational domain of consideration, we include the entire Earth. 
In our graph grammar-based mesh generation procedure, described in Section \ref{sec:CPgraph}, we produce finite element meshes that are defined using spherical coordinates with a fixed radius $R$. 
Hence, classical longitude-latitude coordinates $(
\lambda,\psi)$.
The reference water level corresponds to the sphere. The wave equation seeks a water level $(
\lambda,\psi) \rightarrow u(
\lambda,\psi) \in {\cal R}$. The sea floor is defined by
$(
\lambda,\psi) \rightarrow h_b(
\lambda,\psi) \in {\cal R}$. The three-dimensional mesh obtained by the graph-grammar model approximates the bathymetry, topography, and coastline based on the data provided by \cite{GMRT}.
The generated computational mesh for the whole Earth is presented in Figures \ref{fig:Earth1}-\ref{fig:Earth3}.

To accomplish this, we use standard transformations to this reference frame using the following transformations to a longitude-latitude coordinate system, i.e.,
\begin{equation}
\begin{aligned}
    x&=(\lambda - \lambda_0) \, R \,\text{cos}\,\phi_0, \\    
    y&=\,R \,\phi. 
\end{aligned}
\end{equation}
Here $(\lambda_0,\phi_0)$ are coordinates of a reference point. Using the chain rule, we can perform a substitution and coordinate shift such that: 
\begin{equation} 
   \nabla \left( g(u-h_b) \nabla u \right) = \nabla \{ g(u-h_b)(R\text{cos} \phi_0)^{-1}\frac{\partial u}{\partial \lambda}, g(u-h_b) R^{-1}\frac{\partial u}{\partial \phi} \}^{\text{T}},
\end{equation}
Thus:
\begin{eqnarray} \label{eq:wave_eq_spherical}
\frac{\partial^2 u}{\partial t^2} - R^2 \left( \frac{1}{\text{cos}^2 \phi_0}\frac{\partial}{\partial \lambda} \{g(u-h_b) \frac{\partial u }{\partial \lambda} \}  \right) - R^2 \frac{\partial}{\partial \phi} \left(  g(u-h_b) \frac{\partial u }{\partial \phi}   \right) = 0 \text{ in } \, \Omega.
\end{eqnarray}

\section{Numerical scheme}
\label{sec:genalpha}

\newcommand{\U}{\mathbf{u}}
\newcommand{\A}{\mathbf{a}}
\newcommand{\V}{\mathbf{v}}

To develop numerical simulations using~\eqref{eq:wave_eq_spherical},
we discretize in space using the Finite Element method and use the generalized alpha scheme~\cite{generalized_alpha} to perform time stepping.
After integration by parts and application of boundary conditions, we can write the equivalent weak formulation of~\eqref{eq:wave_eq_spherical} as: 
\begin{equation}
  \Prod{\frac{\partial^2 u}{\partial t^2}}{v} + b(u, v) = 0 \quad \forall v \in V,
\end{equation}
for a bilinear form~$b(\cdot,\cdot)$ and trial space~$V$: 
\begin{eqnarray} \label{eq:wave_eq_spherical2}
b(u,v) =  R^2 g \left(  (u-h_b) \frac{\partial u }{\partial \phi}    , \frac{\partial v}{\partial \phi}  \right) - R^2 g \frac{1}{\text{cos}^2 \phi_0}\left((u-h_b) \frac{\partial u }{\partial \lambda} , \frac{\partial v}{\partial \lambda}\right) \quad \forall v \in V,
\end{eqnarray}
Since we solve the whole Earth, there are no boundary conditions. 
To alleviate encountered stability issues, we also introduce
a non-zero dampening term of the form~$c \frac{\partial u}{\partial t}$, where
\begin{equation}
  c = c_0 \frac{h_0}{\min\{h_0, h_b\}},
\end{equation}
In deep regions of the ocean ($h_b \ll h_0$) we have~$c \approx 0$,
while in the shallow waters ($h_b \approx h_0$) we have~$c \approx c_0$,
meaning that the damping effect is only meaningful near the shores, where
we encountered numerical instabilities.
In the numerical experiments presented in the paper, we used~$c_0 = 5\times 10^{-4}$
and~$h_0 = -5$.

Applying the Bubnov-Galerkin method results in a semi-discrete formulation,
\begin{equation}
  M\ddot{\U} + C \dot{\U} + B \U = 0.
  \label{eq:BG}
\end{equation}
The first term involves the mass matrix, the second term involves the damping matrix, and the last term involves the matrix of our bilinear form.
The solution vector $\U$ contains the coefficients of the discrete solution~$u_h$. We use the generalized-$\alpha$ scheme with time step size $\tau$.
We introduce the acceleration $\A = \ddot{\U}$ as well as we introduce the velocity $\V = \dot{\U}$. Using this notation, we can rewrite our problem (\ref{eq:BG}) into
\begin{equation}
  M\A + C\V + B\U = 0,
\end{equation}
The generalized-$\alpha$ method parameterized with
$\alpha_1$, $\alpha_2$, $\beta$, $\gamma$
involves the following sequence of steps 
\begin{equation}
M \A_{n + 1 - \alpha_1} + C\V_{n + 1 - \alpha_2} + B\U_{n + 1 - \alpha_f} = 0, \\
\end{equation}
where the symbols with fractional indices are defined as
\begin{equation}
\begin{aligned}
\A_{n + 1 - \alpha_1} &= (1 - \alpha_1) \A_{n+1} + \alpha_m \A_n, \\
\V_{n + 1 - \alpha_2} &= (1 - \alpha_2) \V_{n+1} + \alpha_f \V_n, \\
\U_{n + 1 - \alpha_2} &= (1 - \alpha_2) \U_{n+1} + \alpha_f \U_n, \\
\end{aligned}
\end{equation}
and new values of~$\V$, $\U$ are given by
\begin{equation}
\begin{aligned}
\V_{n + 1} &= \V_n + \tau \left[(1 - \gamma)\A_n + \gamma \A_{n+1}\right], \\
\U_{n + 1} &= \U_n + \tau \V_n + \frac{\tau^2}{2}\left[(1 - 2\beta)\A_n +\beta \A_{n+1} \right].
\end{aligned}
\end{equation}
In the numerical tests, their values are
\begin{equation}
\begin{aligned}
  \alpha_1 &= \frac{2\rho - 1}{\rho + 1}, \\
  \alpha_2 &= \frac{\rho}{\rho + 1}, \\
  \beta &= \frac{1}{4} (1 - \alpha_1 - \alpha_2)^2, \\
  \gamma &= \frac{1}{2} - \alpha_1 - \alpha_2, \\
\end{aligned}
\end{equation}
where $\rho = 0.2$. Such the setup of the method parameters makes sure that this time integration scheme is of the second order ~\cite{generalized_alpha}.

\section{Simulation of a tsunami caused by dam break}
\label{sec:break}

We present a numerical simulation of the aftermath of the hypothetical NEED failure, caused by either natural causes, structural failure, or terrorist attack. 
While all these scenarios could cause slightly different initial conditions, we assume that in our initial configuration, the NEED has disappeared, including the Northern and Southern parts, and we perform a simulation of the aftermath.

\begin{figure}
    \centering
    \includegraphics[width=0.5\textwidth]{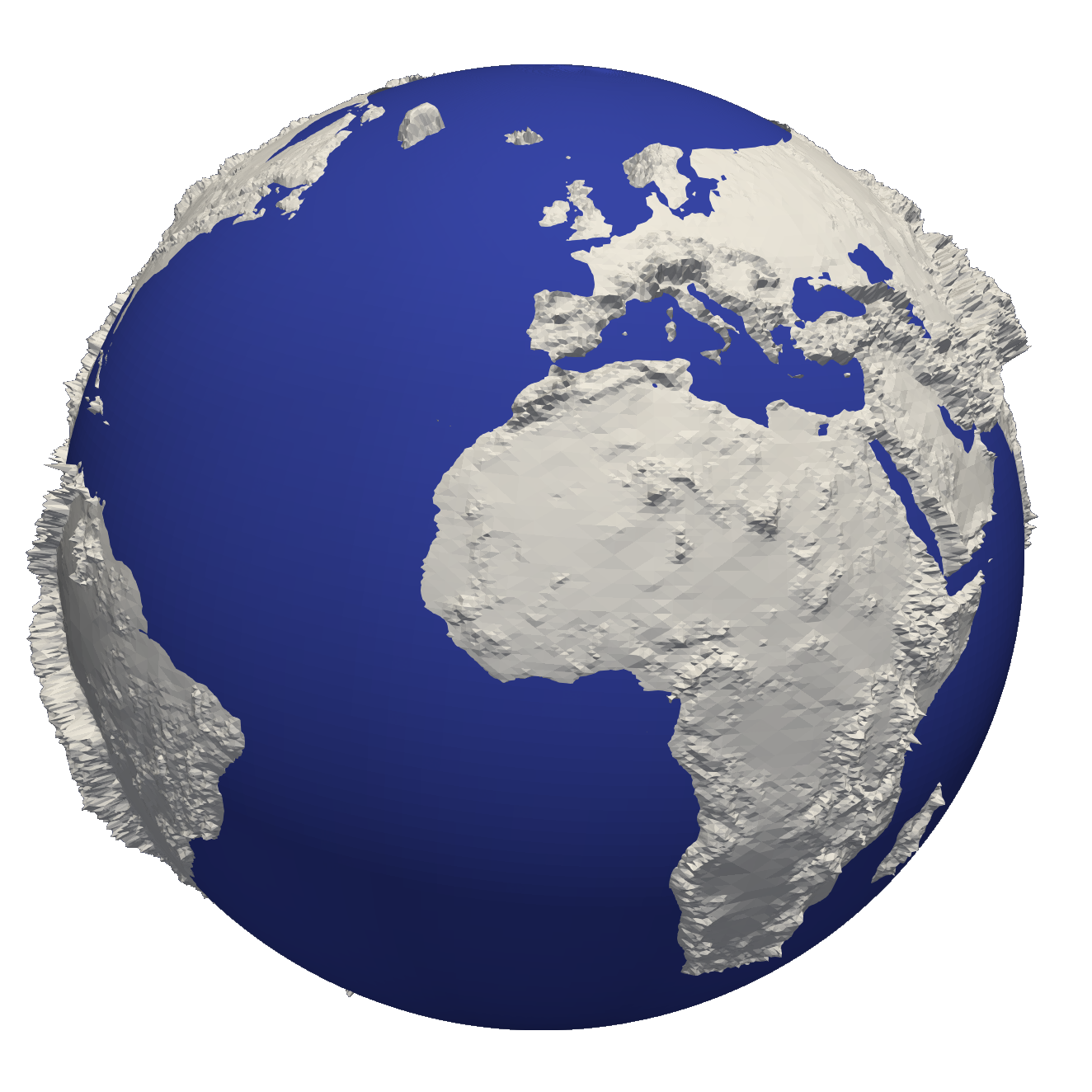}
    \caption{The computational domain of Earth including the coastline, seabed, and the reference sea level. }

    \label{fig:Earth}
\end{figure}

The simulation is performed on a model of the entire Earth, presented in Figure \ref{fig:Earth}, generated using composition graph grammar adapting the coastline and the seabed to the Global Multi-Resolution Topography Data Synthesis database \cite{GMRT}.
We employ a spherical coordinate system since our numerical model describes the ocean water level with respect to a reference level which is shown in blue on the presented sphere.
In our further pictures, we focus on the part of the adaptive mesh describing the Northern Sea, Baltic Sea, and the Northern European Enclosure Dam (NEED).

The simulation performed with our homemade code takes around 2 hours on a laptop having an 11th Generation Intel(R) Core(TM) i5-11500H processor with a 2.92 GHz clock with 32 GB of RAM.
In Section \ref{sec:parallel}, we also summarize the scalability of the parallel code executed on a multi-core node of Ares parallel machine \cite{Ares}.
The simulated time period is 12500 minutes, or 8 days 16 hours and 20 minutes. 

\begin{figure}[!htb]
  \centering
\includegraphics[width=0.7\textwidth]
{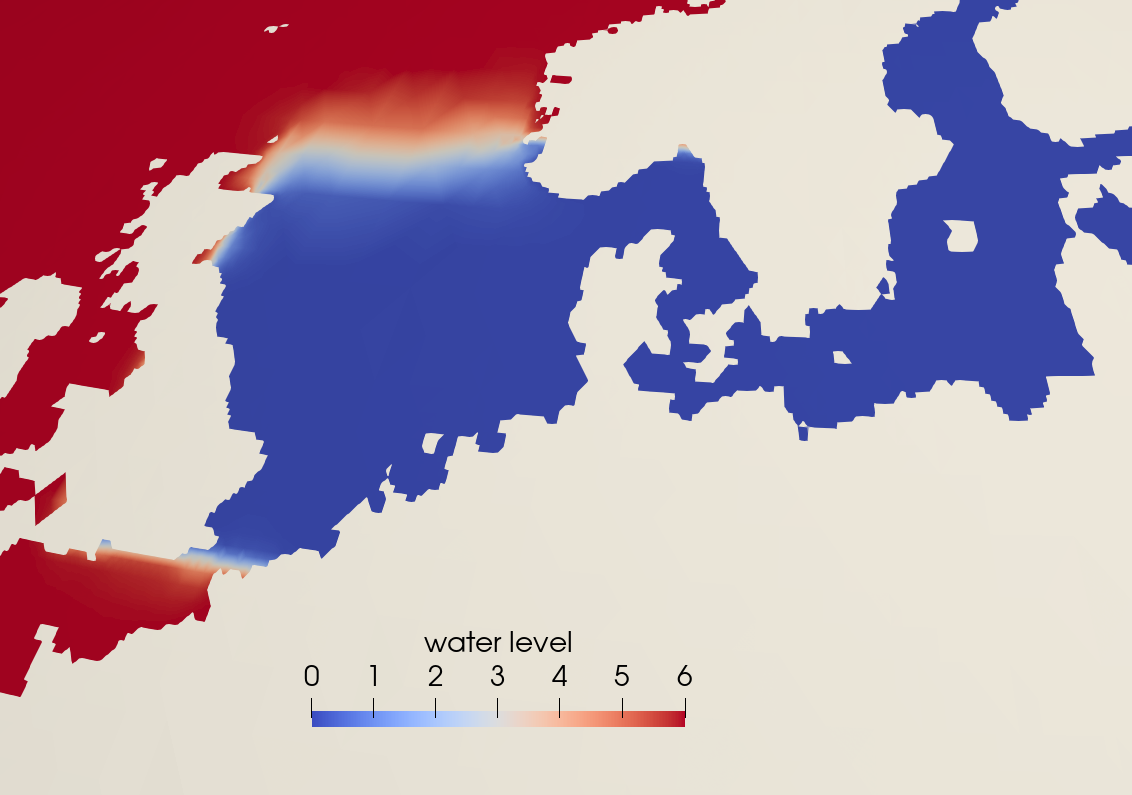}
  \caption{41 minutes after breaking of the NEED.}
  \label{fig:sim1}
\end{figure}

\begin{figure}[!htb]
  \centering
\includegraphics[width=0.7\textwidth]
{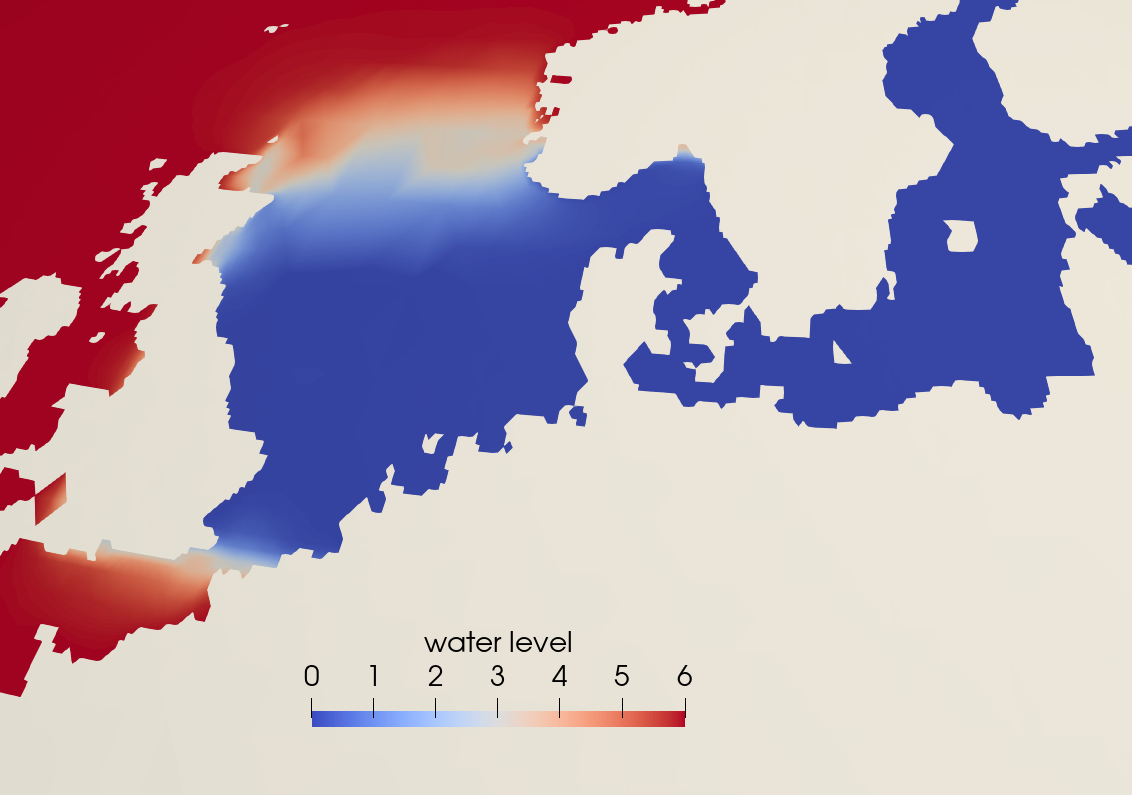}
  \caption{1 hour 23 minutes after breaking of the NEED.}
  \label{fig:sim2}
\end{figure}

\begin{figure}[!htb]
  \centering
\includegraphics[width=0.7\textwidth]
{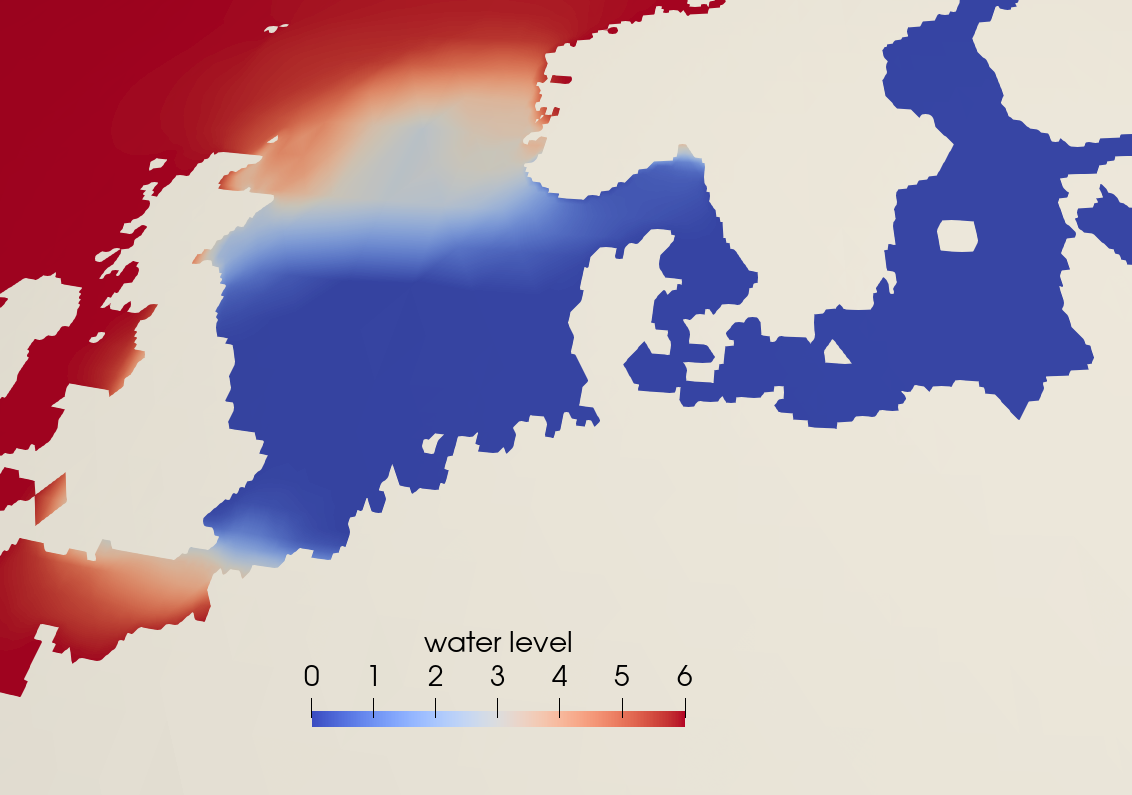}
  \caption{2 hours 5 minutes after breaking of the NEED.}
  \label{fig:sim3}
\end{figure}

\begin{figure}[!htb]
  \centering
\includegraphics[width=0.7\textwidth]
{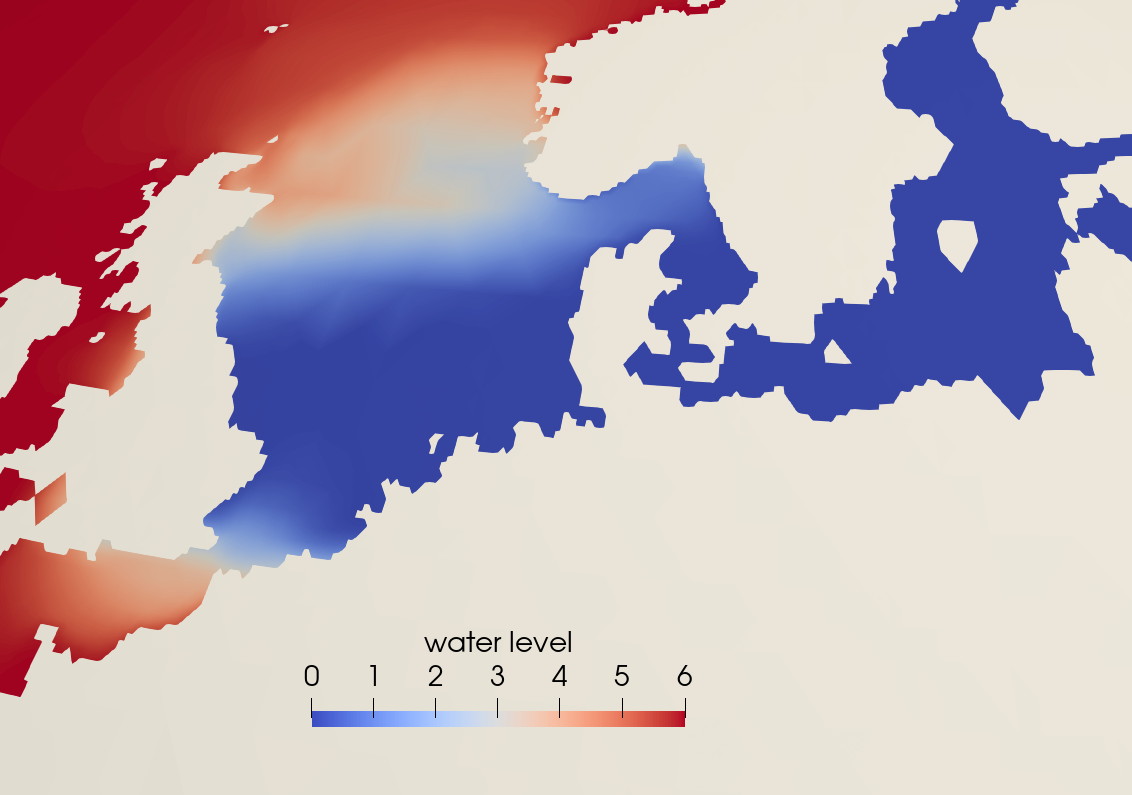}
  \caption{2 hours 46 minutes after breaking of the NEED.}
  \label{fig:sim4}
\end{figure}

\begin{figure}[!htb]
  \centering
\includegraphics[width=0.7\textwidth]
{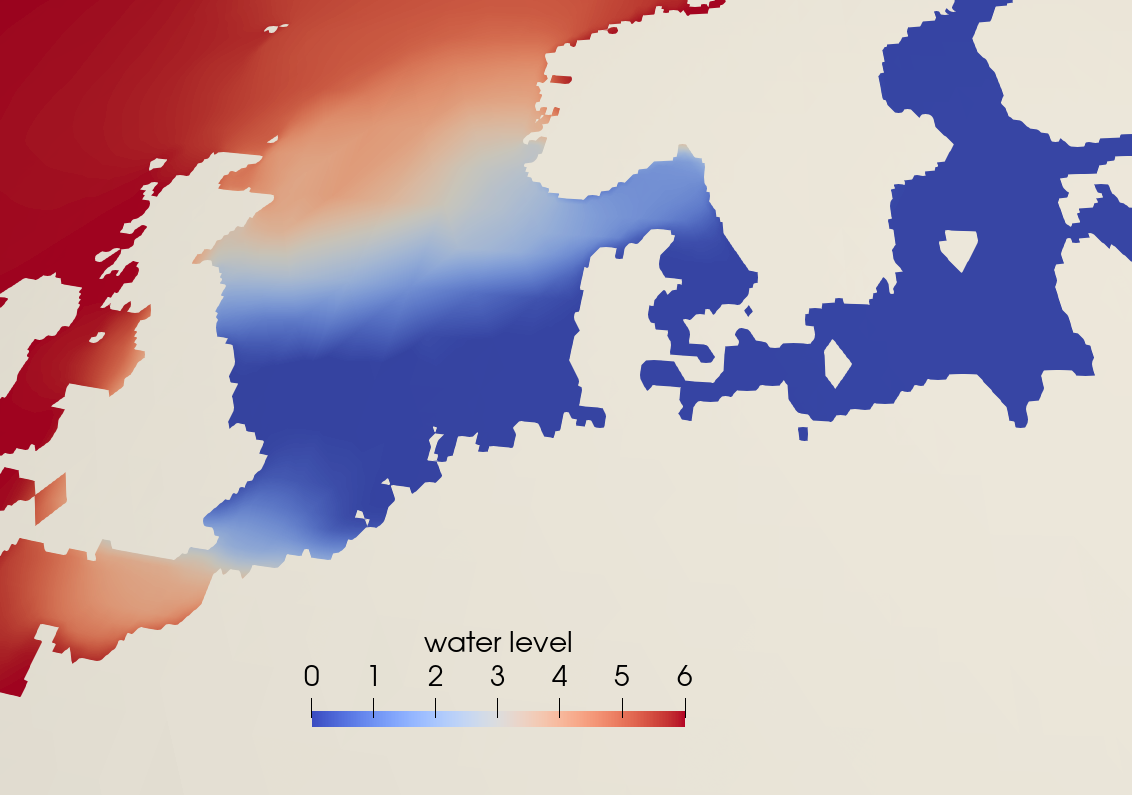}
  \caption{3 hours 28 minutes after breaking of the NEED.}
  \label{fig:sim5}
\end{figure}

\begin{figure}[!htb]
  \centering
\includegraphics[width=0.7\textwidth]
{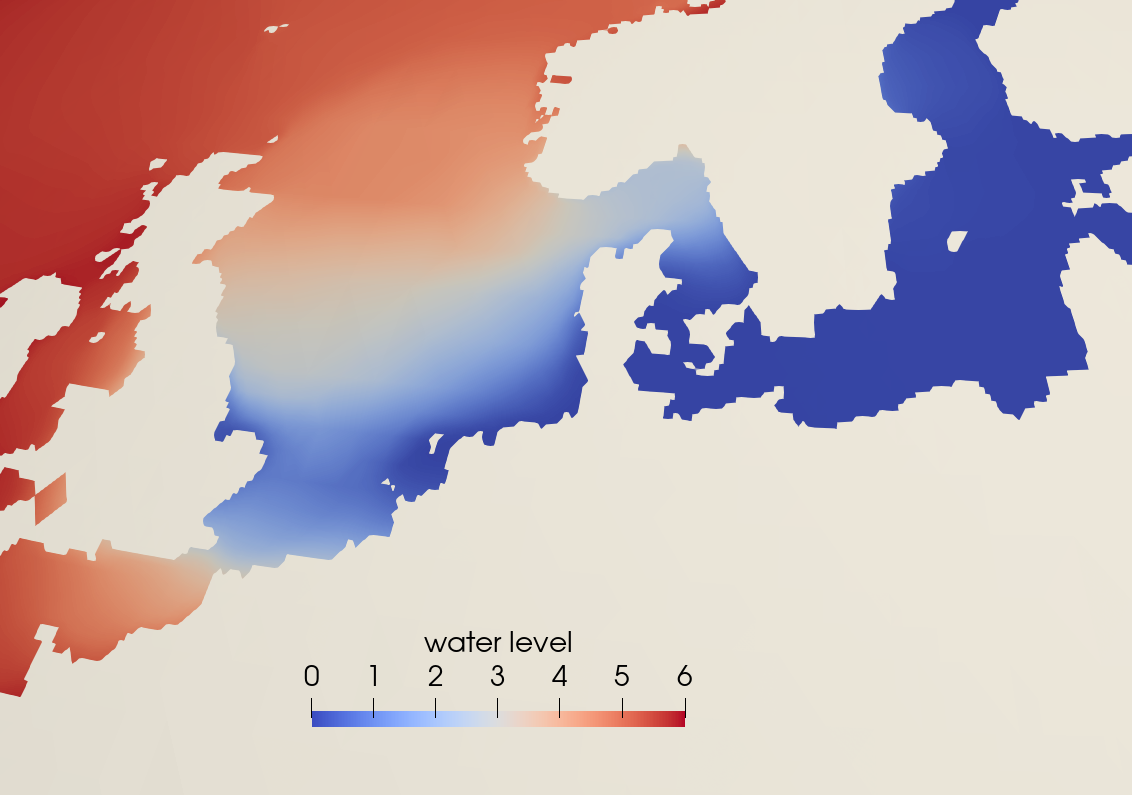}
  \caption{6 hours 56 minutes after breaking of the NEED.}
  \label{fig:sim10}
\end{figure}

\begin{figure}[!htb]
  \centering
\includegraphics[width=0.7\textwidth]
{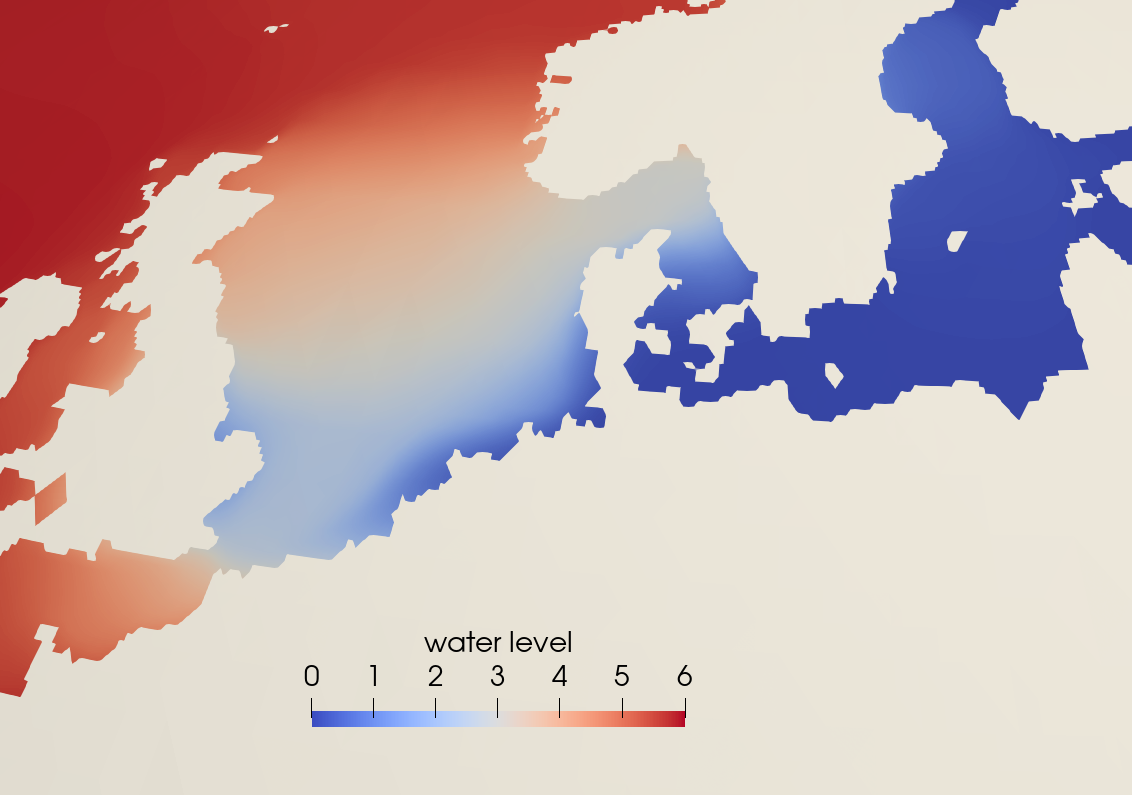}
  \caption{10 hours 25 minutes after breaking of the NEED.}
  \label{fig:sim15}
\end{figure}

\begin{figure}[!htb]
  \centering
\includegraphics[width=0.7\textwidth]
{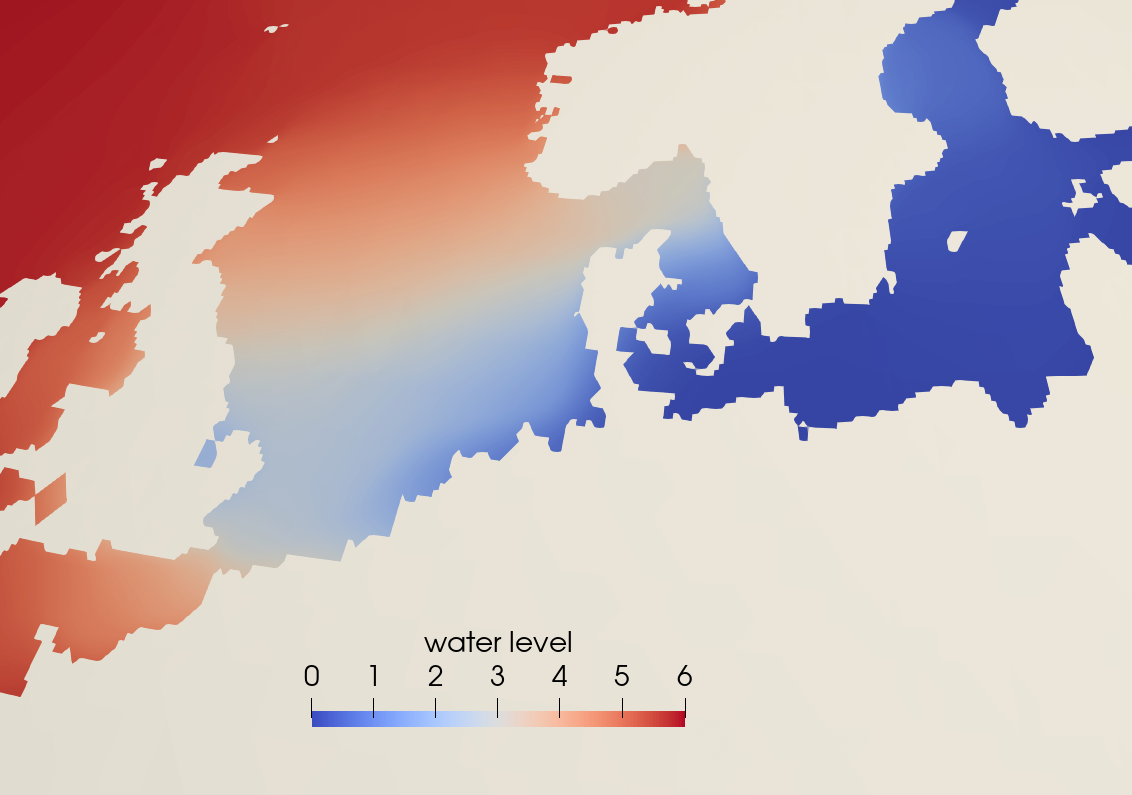}
  \caption{13 hours 3 minutes after breaking of the NEED.}
  \label{fig:sim20}
\end{figure}

\begin{figure}[!htb]
  \centering
\includegraphics[width=0.7\textwidth]
{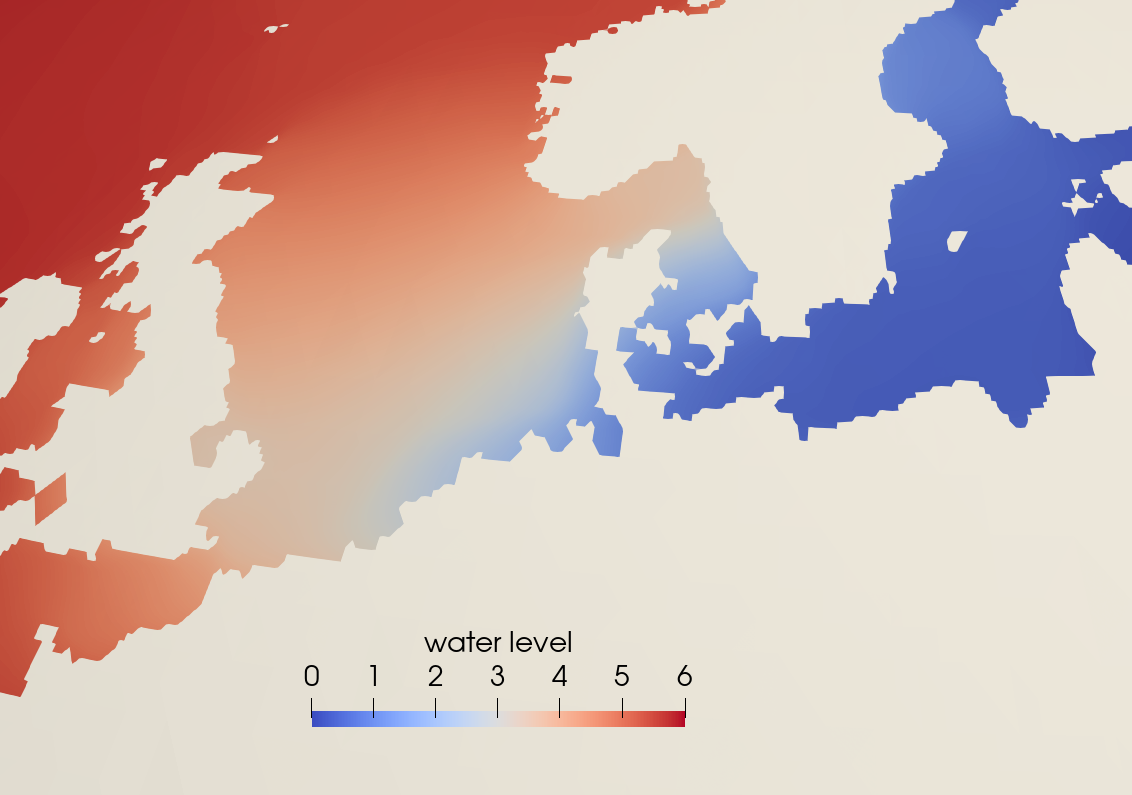}
  \caption{1 day 3 hours 46 minutes after breaking of the NEED.}
  \label{fig:sim40}
\end{figure}

\begin{figure}[!htb]
  \centering
\includegraphics[width=0.7\textwidth]
{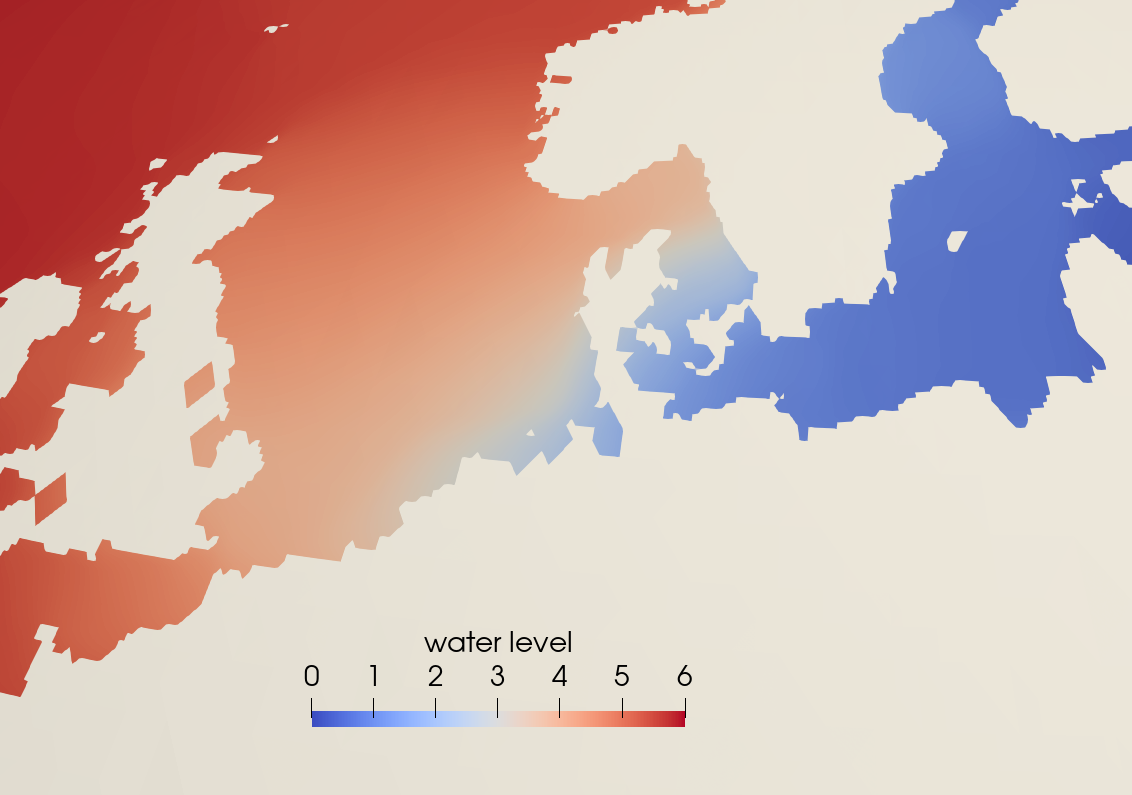}
  \caption{1 day 17 hours 40 minutes after breaking of the NEED.}
  \label{fig:sim60}
\end{figure}

\begin{figure}[!htb]
  \centering
\includegraphics[width=0.7\textwidth]
{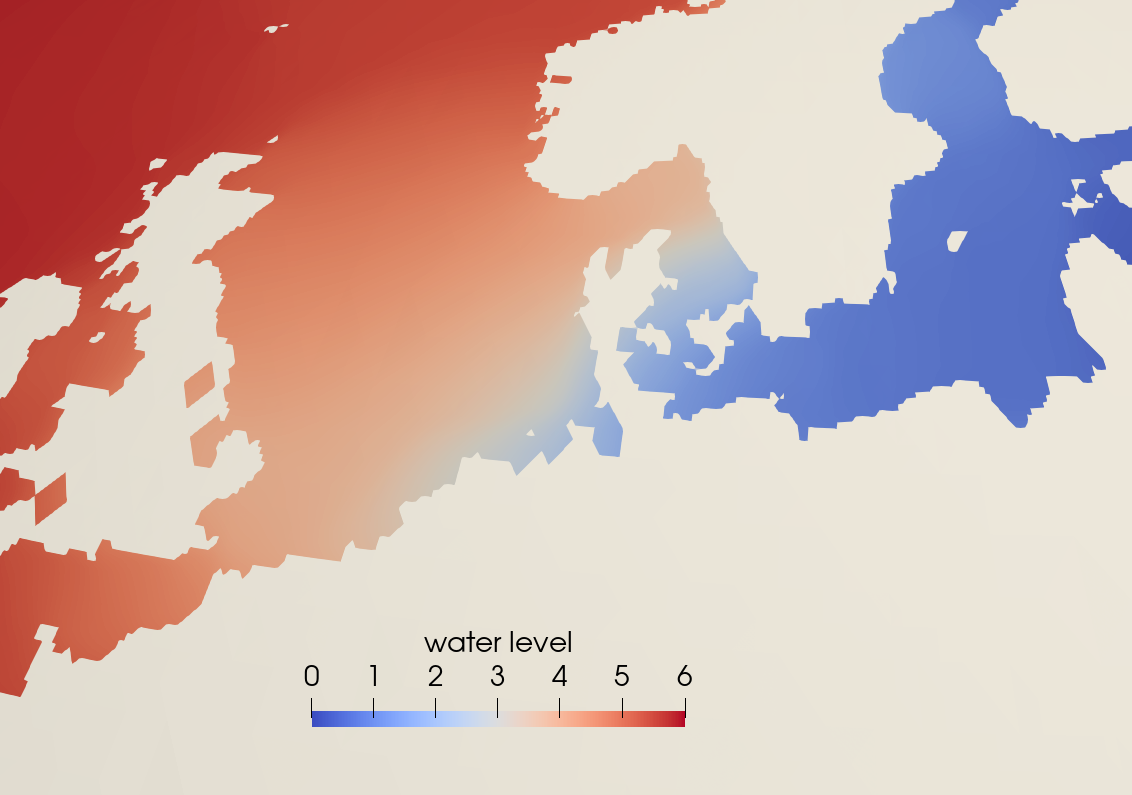}
  \caption{2 days 21 hours 26 minutes after breaking of the NEED.}
  \label{fig:sim100}
\end{figure}

\begin{figure}[!htb]
  \centering
\includegraphics[width=0.7\textwidth]
{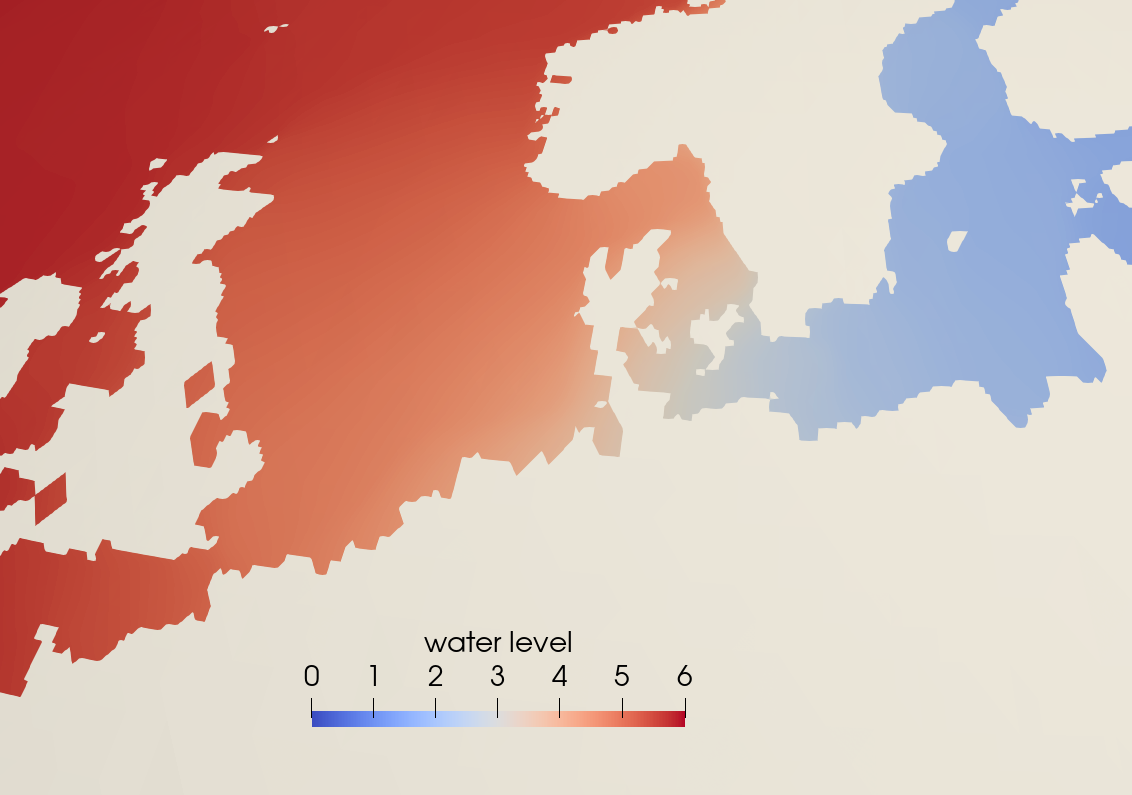}
  \caption{4 days 8 hours 10 minutes after breaking of the NEED.}
  \label{fig:sim150}
\end{figure}

\begin{figure}[!htb]
  \centering
\includegraphics[width=0.7\textwidth]
{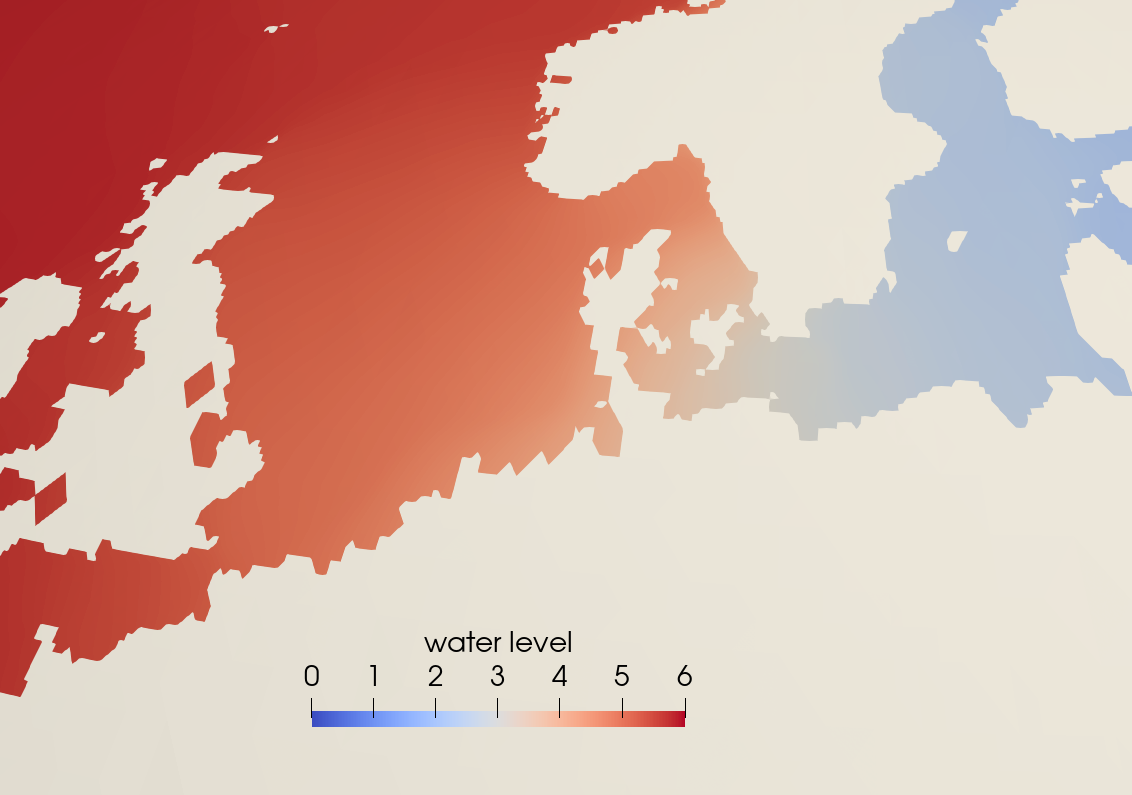}
  \caption{5 days 18 hours 53 minutes after breaking of the NEED.}
  \label{fig:sim200}
\end{figure}

\begin{figure}[!htb]
  \centering
\includegraphics[width=0.7\textwidth]
{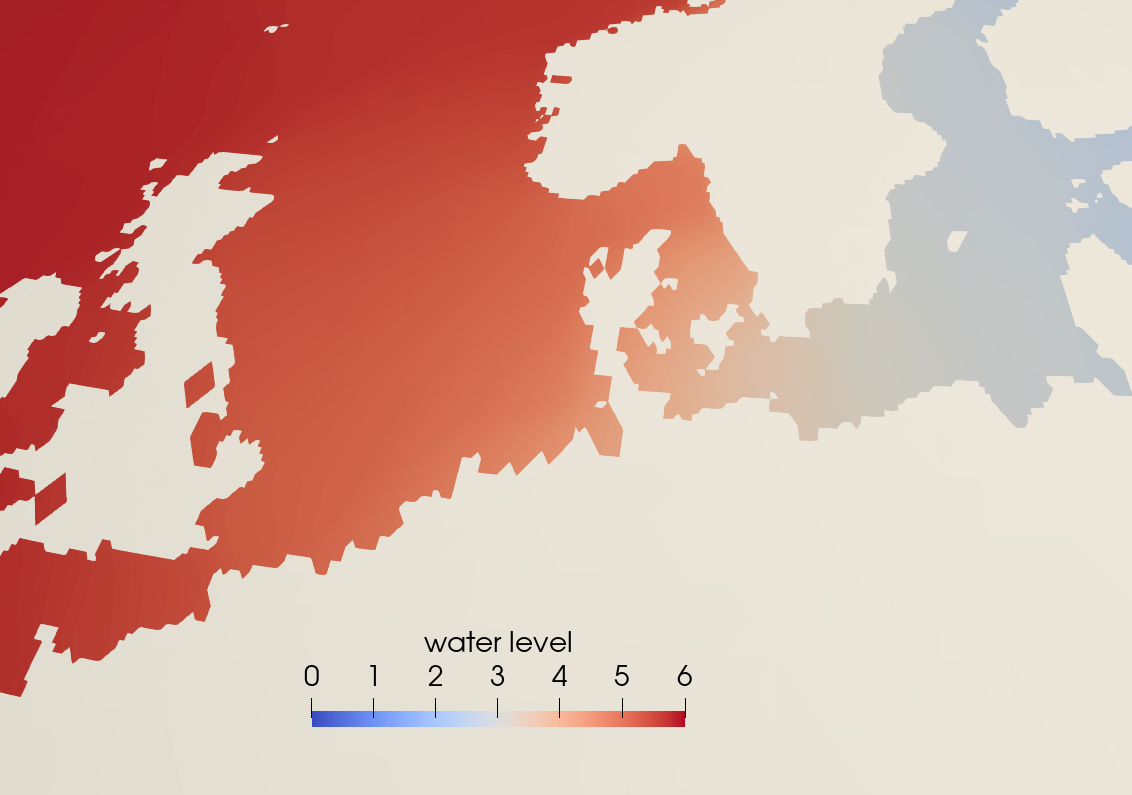}
  \caption{7 days 5 hours 36 minutes after breaking of the NEED.}
  \label{fig:sim250}
\end{figure}

\begin{figure}[!htb]
  \centering
\includegraphics[width=0.7\textwidth]
{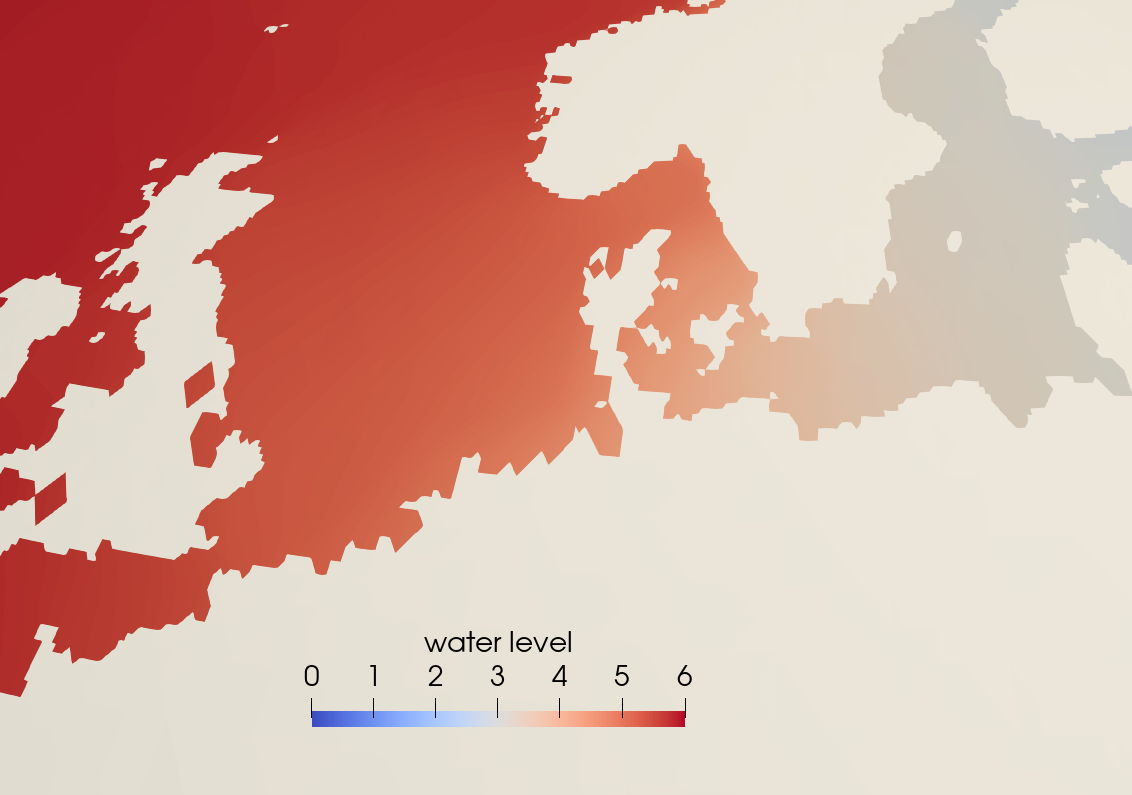}
  \caption{8 days 16 hours 20 minutes after breaking of the NEED.}
  \label{fig:sim300}
\end{figure}

Subsequent snapshots from the computer simulation, summarized in Table \ref{tab:tab1} are shown in Figures \ref{fig:sim1}-\ref{fig:sim300}. 
\begin{table}[]
    \centering
    \begin{tabular}{c|c|c}
\hline
Figure & time step & simulated time \\
\hline
18 & 1 & 41 minutes \\
19 & 2 & 1 hours 23 minutes \\
20 & 3 & 2 hours 5 minutes \\
21 & 4 & 2 hours 46 minutes \\
22 & 5 & 3 hours 28 minutes \\
23 & 10 & 6 hours 56 minutes \\
24 & 15 & 10 hours 25 minutes \\
25 & 20 & 13  hours 3 minutes \\
26 & 40 & 1 day 3 hours 46 minutes \\
27 & 60 & 1 day 17 hours 40 minutes \\
28 & 100 & 2 days 21 hours 26 minutes \\
29 & 150 & 4 days 8  hours 10 minutes \\
30 & 200 & 5 days 18 hours 53 minutes \\
31 & 250 & 7 days 5 hours  36 minutes \\
32 & 300 &  8 days 16 hours 20 minutes \\
\hline
    \end{tabular}
    \caption{Snapshots of the simulation}
    \label{tab:tab1}
\end{table}
These figures show the systematic flooding of the North Sea basin and then the Baltic Sea by a diffusing wave with a speed of about 10 kilometers per hour. In this stage, the level of the North Sea and the Baltic Sea gradually rises to the initially assumed level of 6 meters above the original average sea level.

\section{Parallel computations}
\label{sec:parallel}

\begin{figure}[!htb]
  \centering
\includegraphics[width=0.7\textwidth]
{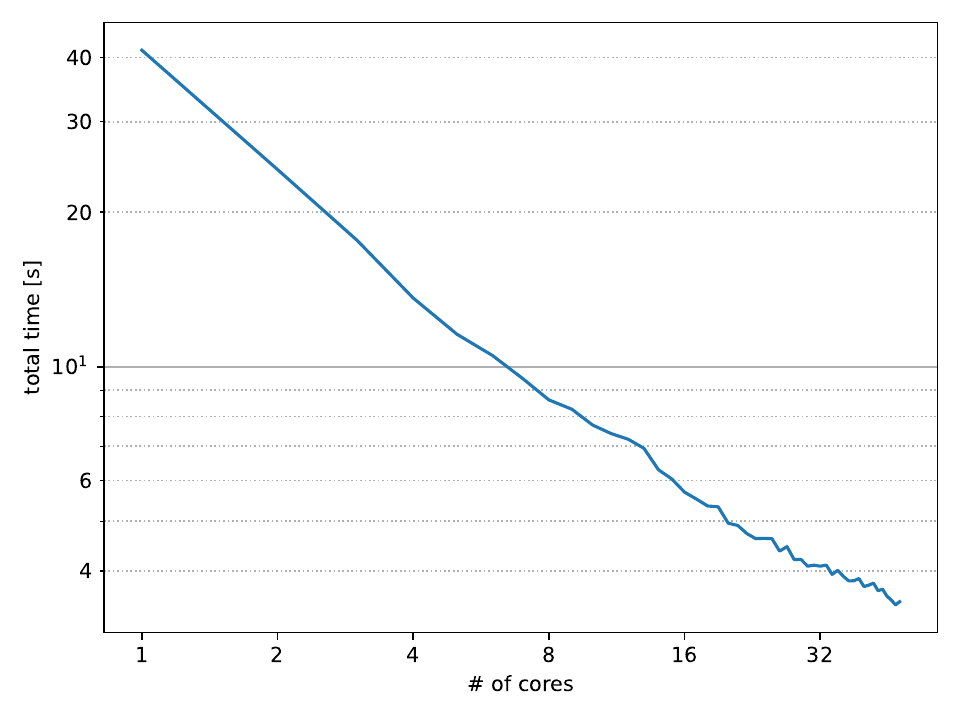}
  \caption{Execution time of parallel code on a single node of ARES with 48 cores and 192 GB of RAM.}
  \label{fig:time}
\end{figure}

\begin{figure}[!htb]
  \centering
\includegraphics[width=0.7\textwidth]
{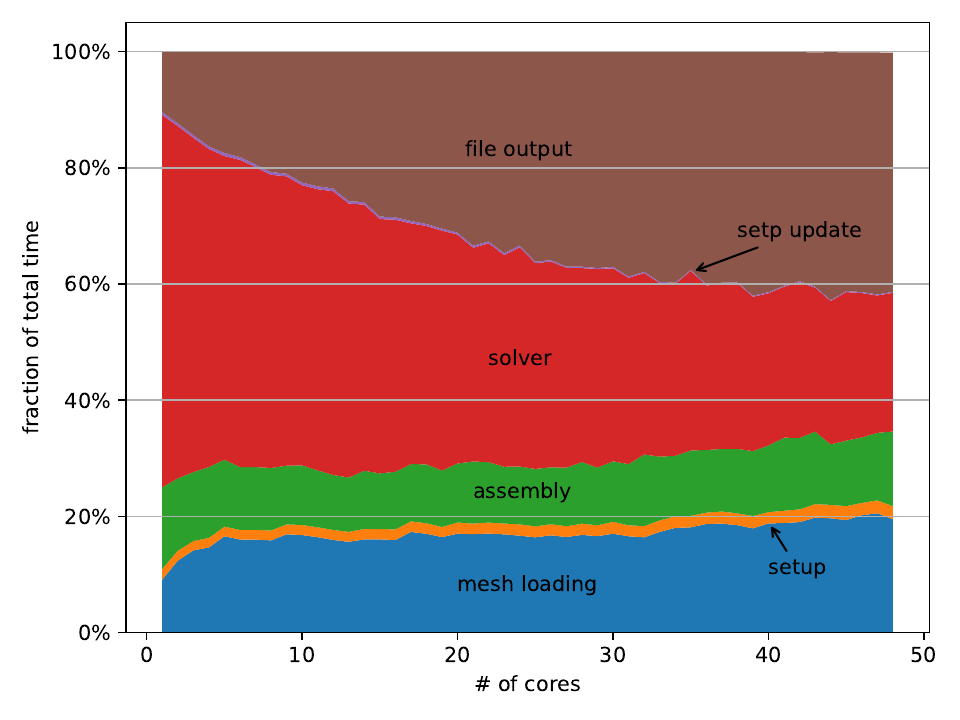}
  \caption{Fractions of the execution times of particular parts of the  parallel code executed on a single node of ARES with 48 cores and 192 GB of RAM.}
  \label{fig:fraction}
\end{figure}

We have run 20 time iterations of our simulator on a single node of ARES supercomputer \cite{Ares} equipped with 48 cores and 192 GB of RAM, measuring:
\begin{itemize}
\item "mesh loading" - mesh file loading time,
\item "setup" - time to allocate memory, pre-process the mesh, initial state, etc.
\item "assembly" - time spent generating matrices and right sides for the system of equations,
\item "solver" - time of linear solver,
\item "step update" - the updates of the velocity and wave shape fields, estimated from the accelerations solved,
\item "file output" - saving the results to a file.
\end{itemize}
The execution times of different parts of the algorithm, 
measured as we increase the number of cores, from 1 to 48 cores, are presented in Figure \ref{fig:time}.
The fractional decompositions of the total simulation time into different parts for increasing number of cores are outlined in Figure \ref{fig:fraction}.

\begin{figure}[!htb]
  \centering
\includegraphics[width=0.7\textwidth]
{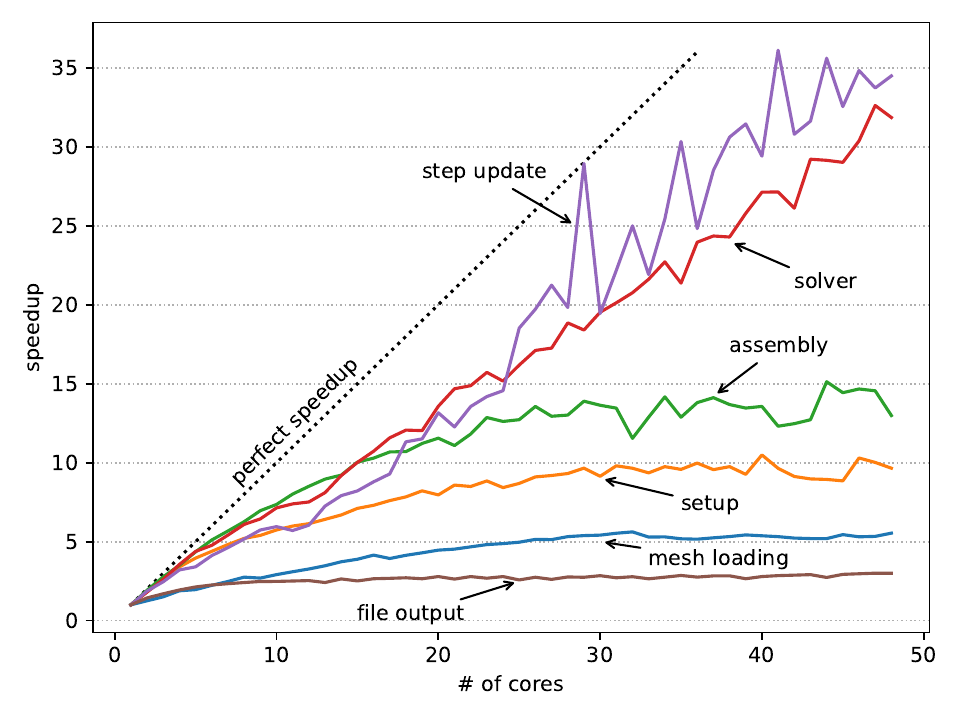}
  \caption{Speedup of parallel execution on a single node of ARES with 48 cores and 192 GB of RAM.}
  \label{fig:speedup}
\end{figure}

\begin{figure}[!htb]
  \centering
\includegraphics[width=0.7\textwidth]
{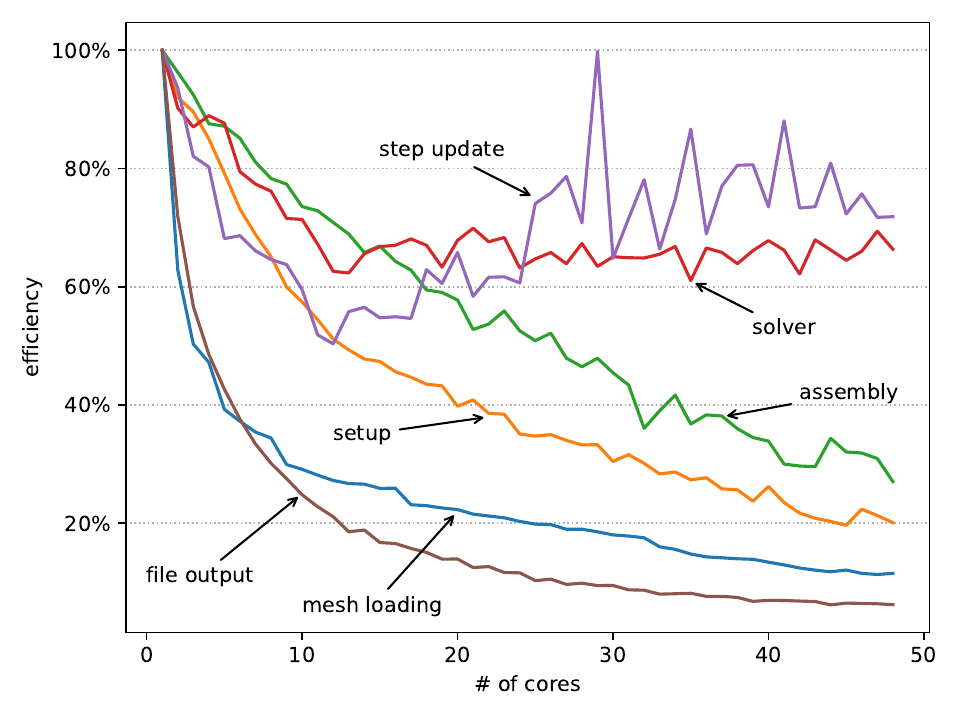}
  \caption{Efficiency of parallel execution on a single node of ARES with 48 cores and 192 GB of RAM.}
  \label{fig:efficiency}
\end{figure}

The summary of the speedup of the particular parts, measured for 1-48 cores, is presented in Figure \ref{fig:speedup}.
The efficiency for 1-48 cores is presented in Figure \ref{fig:efficiency}.
To summarize the execution times, we can perform 20 iterations of the simulator using 48 cores in less than 4 seconds; see Figure \ref{fig:time}.

\section{Conclusions}
\label{sec:conclusions}

The massive damage to the Northern European Enclosure Dam is predicted to have huge consequences for the European population living in coastal areas. 

We combined a unique graph transformation model to generate the entire Earth to perform a simulation of the 
NEED break using wave equation, finite element method solver with the higher-order generalized-$\alpha$ method.

From our simulations, we can draw the following conclusions: a potential NEED should be built with great care and certainly with additional coastal protection systems in the event of a catastrophic break. 

\section{Acknowledgements}
The authors are grateful for support from the funds the Polish Ministry of Science and Higher Education assigned to AGH University of Krakow.
The work supported by  ``Excellence initiative - research university" for the AGH University of Krakow.
The work of Albert Oliver-Serra was supported by 
"Ayudas para la recualificación del sistema universitario español" grant funded by the ULPGC, the Ministry of Universities by Order UNI/501/2021 of 26 May, and the European Union-Next Generation EU Funds.

\section{Computer Code Availability}
NEED dam break simulation was carried out using tsunami-europe script, developed by Marcin \L{}o\'s, using FEniCS framework 

\url{https://fenicsproject.org/download/archive/}

Running it requires Python (version at least 3.10), FEniCS framework and meshio Python package. No special hardware is required. The script consists of 170 lines of Python code. The source code and the mesh file are freely available at

\url{https://github.com/marcinlos/tsunami-europe}

under MIT License, with a brief instruction.

The meshes used in the NEED dam break simulation have been generated
using the tsunami-europe-mesh script developed by Pawe\l{} Maczuga and
Albert Oliver-Serra. You can access the script in the following
publicly available repository:

\url{https://github.com/albert-oliver/Tsunami-Europe-mesh}

The code runs in the Julia programming language, and it depends on the
following two libraries, also developed by Pawe\l{} Maczuga and Albert Oliver-Serra:

\url{https://github.com/albert-oliver/MeshGraphs.jl}

This library implements
the Graph Grammar for the Rivara refinement and provides an API for
generating and refining a mesh.

\url{https://github.com/albert-oliver/TerrainGraphs.jl}

This library
implements the mesh generation of terrains (either flat or spherical).
It reads a GeoTIFF file, generates an initial coarse mesh, checks the
refinement criteria specific to terrains, and refines the mesh until
it obtains the final mesh.

Please note that the software code is available under the MIT license

\bibliographystyle{splncs04}

\end{document}